\documentclass{aa}

\usepackage{longtable,lscape}
\usepackage{natbib}
\bibpunct{(}{)}{;}{a}{}{,} 

\usepackage{graphicx}

\def\cd  {{$\mbox{c~d}^{-1}$}}
\defcitealias{gutierrez2007}{GS07}
\defcitealias{2007A&A...462..683M}{M07}

\title{Pulsating B and Be stars in the Small Magellanic Cloud\thanks{Figs.~\ref{PlotFases-B},~\ref{PlotFases-Be-4},~\ref{PlotBeatingTots}-\ref{OutCompo} and Tables~\ref{BeatingTable}-\ref{OutTab}, are only available in electronic form at {\tt http://www.aanda.org}}}
\author{P. D. Diago\inst{1}
\and J. Guti\'{e}rrez-Soto\inst{1,2} \and
J. Fabregat\inst{1,2} \and C. Martayan \inst{2,3}}
\offprints{Pascual.Diago@uv.es}
\institute{Observatori Astron\`{o}mic de la Universitat de Val\`{e}ncia, Edifici Instituts d'Investigaci\'{o}, Pol\'{i}gon La Coma, 46980 Paterna, Val\`{e}ncia, Spain
\and 
GEPI, Observatoire de Paris, CNRS, Universit\'e Paris Diderot; place Jules Janssen 92195 Meudon Cedex, France
\and
Royal Observatory of Belgium, 3 Avenue Circulaire, B-1180 Brussels, Belgium
}

\date{Received / Accepted }

\abstract
{Stellar pulsations in main-sequence B-type stars are driven by the $\kappa$-mechanism due to the Fe-group opacity 
bump. The current models do not predict the presence of instability strips in the B spectral domain at 
very low metallicities. As the metallicity of the SMC is lower than $Z=0.005$, it constitutes 
a very suitable object to test these predictions.
}
{The main objective is to investigate the existence of B-type pulsators at low metallicities, searching 
for short-term periodic variability in absorption-line B and Be stars in the SMC. 
The analysis has been performed in a sample of 313 B and Be stars with fundamental 
astrophysical parameters accurately determined from high-resolution spectroscopy.}
{Photometric light curves of the MACHO project have been analyzed using standard Fourier techniques and linear and non-linear least squares fitting methods. The position of the pulsating stars in the HR diagram has been 
used to ascertain their nature and to map the instability regions in the SMC.}
{We have detected 9 absorption-line B stars showing short-period variability, two among them being multiperiodic. 
One star is most likely a $\beta$ Cephei variable and the remaining 8 are SPB stars. The SPB instability strip in the SMC is shifted towards higher temperatures than the Galaxy. In the Be star sample, 32 stars are short-period
variables, 20 among them multiperiodic. 4.9\% of B stars and 25.3\% of Be stars are pulsating stars. 
}
{
$\beta$ Cephei and SPB stars do exist at the SMC metallicity. The fractions of SPB stars and pulsating Be stars in the SMC are
lower than in the Galaxy. The fraction of pulsating Be stars in the SMC is much 
higher than the fraction of pulsating absorption-line B stars, as in the Galaxy.
}
\keywords{Stars: emission-line, Be -- Stars: oscillations (including pulsations) -- Stars: early-type -- Stars: statistics -- Galaxies: Magellanic Clouds}

\begin{document}
\maketitle

\section{Introduction}

A significant fraction of main-sequence B-type stars are known to be variable. The whole main-sequence in the B spectral range is populated by stars belonging to two well characterized classes of pulsating variables: the $\beta$ Cephei stars \citep{2005ApJS..158..193S} and the Slowly Pulsating B stars \citep[SPB stars,][]{2002ASPC..259..196D}. Both types of stars pulsate due to the $\kappa$-mechanism acting in the partial ionization zones of iron-group elements. $\beta$ Cephei stars do pulsate in low-order p- and g-modes with periods similar to the fundamental radial mode. SPB stars are high-radial order g-mode pulsators with periods longer than the fundamental radial one.

The $\kappa$-mechanism in $\beta$ Cephei and SPB stars has an important dependence on the abundance of iron-group elements, and hence the respective instability strips have a great dependence on the metallicity of the stellar environment. \citet{1999AcA....49..119P} showed that the $\beta$ Cephei and SPB instability strips practically vanish at $Z < 0.01$ and $Z < 0.006$, respectively. 
The metallicity of the Small Magellanic Cloud (SMC) has been measured to be between $Z = 0.001$ and $Z = 0.004$ \citep[see][and references therein]{1999A&A...346..459M}. 
Therefore, it is not expected to find $\beta$ Cephei or SPB pulsators in the SMC. However, as observational capabilities progress, B-type pulsators are found in low-metallicity environments \citep[see, e.g.,][and references therein]{2006MmSAI..77..336K}. Recently, \citet{2007MNRAS.375L..21M,2007CoAst.151...48M} have made new calculations based on OPAL and updated OP opacities and different metal mixtures. They have shown that the blue border of the SPB instability strip is displaced at higher effective temperatures at solar metallicity, and that a SPB instability strip exists at metallicities as low as $Z=0.005$. Their calculations, however, do not predict $\beta$ Cephei pulsations at $Z = 0.005$.

Another large class of stars populating the B-type main sequence are the Be stars. They are defined as  
non-supergiant B star whose spectrum has or had at some time one or more Balmer lines in emission. As the Be nature is transient, Be stars might exhibit a normal B-type spectrum at times, and hitherto normal B stars may become Be stars. Classical Be stars are physically understood as rapidly rotating B-type stars with line emission arising from a circumstellar disk in the equatorial plane, made by matter ejected from the stellar photosphere by mechanisms not yet understood \citep[see][for a recent review]{2003PASP..115.1153P}.

Be stars show two different types of photometric variability with different origin and time-scale: (i) Long-term variability due to variations in size and density of the circumstellar envelope. Variations are irregular and sometimes quasi-periodic, with time scales of weeks to years; (ii) Short-term periodic variability, with time-scales from 0.1 to 3 days, attributed to the presence of non-radial pulsations. 
As pulsating Be stars occupy the same region of the HR diagram as $\beta$ Cephei and SPB stars, it is generally assumed that pulsations have the same origin, i.e., p and/or g-mode pulsations driven by the $\kappa$-mechanism associated with the Fe bump. 
The current theoretical models are not suitable to describe the pulsational characteristics of Be stars, 
due to the high rotational velocity of these objects.

However, recent developments 
are contributing to progress on this issue \citep[e.g.][]{reese06}.
In a recent study of the pulsational properties of Be stars in the Galaxy, \citet[][hereafter GS07]{Gutierrez-Soto2007} have shown that the fraction of pulsators among Be stars is much higher than in absorption-line slow-rotating stars in the same spectral range.

The aim of this work is to search for short-term periodic variability in both B and Be stars in the SMC. We have focused our research on a sample of more than 300 B-type stars for which \citet[][hereafter M07]{2007A&A...462..683M} obtained high-resolution spectroscopy with the FLAMES instrument at ESO/VLT. M07 have identified the Be stars,
provided spectral classification for the whole sample and accurately calculated the fundamental astrophysical parameters, including $\log T_{\mathrm{eff}}$ and $\log L/L_{\odot}$. This will allow us to place the periodic variables in the theoretical HR diagram in order to better understand their nature and to map the regions of pulsational instability at the SMC metallicity. A preliminary discussion of some of the pulsating Be stars found in this work has already been presented by \citet{Martayan2007}.

The search for periodic variability has been done by analyzing the photometric time series provided by the MACHO 
project \citep{1999PASP..111.1539A}. MACHO is a microlensing survey experiment that monitored about 60 million stars in the Magellanic Clouds and the Galactic Bulge. 
The time span of the MACHO observations is large enough to provide a very high frequency resolution in the spectral analysis and hence allow us to distinguish between very close frequencies.

\section{Data analysis}

\subsection{The sample}

A total of 198 absorption-line B stars were studied by M07. Among them, 4 are not included in the MACHO database and 8 more do not have measured errors in their photometry. From the remaining 186 stars, we consider 3 to be Be stars, since they show outbursts in their light curves (SMC5\_074305 and SMC5\_079021) or long-term quasi-periodic variations (SMC5\_082379). These characteristics are typical of Be stars. The transient nature of the Be phenomenon can explain the non-detection by M07 of the H$\alpha$ emission in these 3 stars. Therefore, the final sample of absorption-line B stars for which we have performed frequency analysis is composed of 183 stars. 

The initial sample of Be stars studied by M07 is composed of 131 objects. 2 of them are not present in the MACHO database, and 2 more do not have measured errors. On the other hand, we have added the 3 stars considered to be Be stars as explained in the above paragraph. Therefore, our sample of Be stars with MACHO photometry consists of 130 objects. Among them, 4 present a very complex long-term light curve, with bursting behavior and irregular high-amplitude variations, which prevents the search for short-term variability. The final Be star sample for which we have performed frequency analysis is thus composed of 126 stars. 

\begin{figure}
\resizebox{\hsize}{!}{\includegraphics{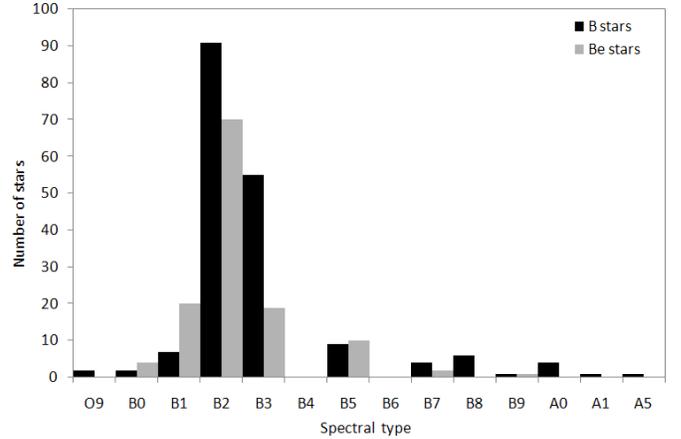}}
\caption{Distribution of our sample of B and Be stars from the SMC as a function of their spectral types.}
\label{HistoSample}
\end{figure}

The distribution of our definitive samples of B and Be stars as a function of their spectral type is presented in Fig.~\ref{HistoSample}. The spectral classification has been taken from M07.

\subsection{Frequency analysis}

The MACHO survey provides photometric instrumental magnitudes for each star in 
two contiguous ``blue'' and ``red'' passbands, which are labeled B and R. They are at different wavelengths than the standard B and R passbands in the Johnson-Cousins photometric system.
The light curves have an average of 1000 data points spanning 2700 days. A typical 1-day alias pattern is present in the spectral window of the MACHO data, since the observations were obtained at only one site.

The frequency analysis was performed in both B and R datasets using a self-developed code based on standard Fourier techniques, and linear and non-linear least-square fitting. This code is quasi-automatic and suitable for medium to large sets of data.

The first step of our method consists of searching for frequencies one by one by computing the discrete Fourier transform \citep{1975Ap&SS..36..137D, 1982ApJ...263..835S} of the light curve, and then adjusting a sinusoidal function for the detected frequency by means of linear least-square fitting. When a frequency is found, it is prewhitened from the original data and in a new step a search begins for a new frequency in the residuals.
The method is iterative and stops when the new frequency is not statistically significant.

The criterion used to determine whether the frequencies are statistically significant is the \emph{signal to noise amplitude ratio requirement} described in \citet{1993A&A...271..482B}. It consists of the calculation of the SNR of the frequency in the periodogram. This is made by calculating the signal as the amplitude of the peak for the frequency obtained and the noise as the average amplitude in the residual periodogram after the prewhitening of all the frequencies detected. \citet{1993A&A...271..482B} proposed that a value of SNR $\geq4$ is a reliable criterion to distinguish between peaks due to real frequencies and noise.

Once we have obtained a preliminary set of significant frequencies for each star and filter, we perform a non-linear multi-parameter fitting based on the method described by \citet{1971Ap&SS..12...10V}, also explained in detail by \citet{1997MNRAS.290..401Z}. This code simultaneously adjusts all the frequencies, allowing the fit to move over a wide range in frequency (global fitting) in order to obtain the minimum variance.

Finally, we determine the amplitudes and phases for the definitive frequencies by least-square fitting to the data. The SNR are calculated for the new set of frequencies. Only significant frequencies are retained from the spectral analysis. The phase diagrams folded with the detected frequencies are displayed in Figs.~\ref{PlotFases-B} to~\ref{PlotFases-Be-4}, available in the electronic version.

The resolution in frequency for the majority of stars is $\sim 3.7 \times 10^{-4}$, following the Rayleigh criterion. The uncertainty on the detected frequencies
is $1 - 5 \times 10^{-5}$~\cd. This value has been derived analytically using the formula given by \citet{1999DSN-M&O'D}, and taking into account the correlations in the residuals, as described by \citet{1991MNRAS.253..198S}.

\section{Results}

In this section we present the results of the search for variability performed on the two samples of B and Be stars  described above.  A summary of these results is shown in Table~\ref{Results}.

\begin{table}
\caption{Results of the variability search in the B and Be star samples.}
\label{Results}
\centering
\begin{tabular}{l c c }
\hline\hline
			& B stars	& Be stars	\\
\hline
Number of stars		& 183		& 130		\\
Short-period variables		& 9		& 32		\\
\ \ \ \ \ - single period	& 7		& 12		\\
\ \ \ \ \ - multiple periods	& 2		& 20		\\
Long-period variables 		& 0		& 6		\\
Eclipsing Binaries		& 3		& 1		\\
Outburst/Irregular variability	& 0		& 5 		\\ 
\hline
\end{tabular}
\end{table}


\subsection{Short-term variability}

We consider as short-period variables those stars whose light curves present periodic variations with 
frequencies higher than 0.333~\cd. In the sample of 183 absorption-line B stars we have found 9 short-period variables. Two among them are multiperiodic. In the sample of Be stars, 126 have been analyzed for short-term variability. 32 Be stars have been found to be short-period variables, 20 among them being multiperiodic.

The detected frequencies with their amplitudes and phases for B and Be stars are presented in Tables~\ref{B-SP} and~\ref{Be-SP} respectively. The phase diagrams for each star folded with the detected frequencies are depicted in Figs.~\ref{PlotFases-B} and~\ref{PlotFases-Be-1} for B and Be stars respectively (only available in the electronic version).

\begin{table*}
\caption{Pulsational characteristics of short-period variable absorption-line B stars in the SMC.}
\label{B-SP}
\centering
\begin{tabular}{ c @{\ \ } c @{\ \ } c @{\ \ } c @{\ \ } c @ {\ \ } c @{\ \ } c @{\ \ \ \ } c @{\ \ } c @{\ \ } c }
\hline
\hline
\multicolumn{2}{ c}{Star iD}&&& \multicolumn{3}{c}{B filter}  & \multicolumn{3}{c }{R filter} \\
EIS iD         & MACHO ID       & Sp. T. & $V\!\sin i$& Freq		& Amp	   & Phase & Freq	& Amp	   & Phase \\ 
               &                &        & [km~s$^{-1}$]     & [\cd]		& [mmag]   & [rad] & [\cd]	& [mmag]   & [rad] \\
\hline

SMC5\_004326	& 207.16318.77	& B1V   &373        &3.63947 	& 7.65    & 4.19 & 3.63950 	& 9.23    & 4.09 \\
		&		&       &        &3.68789 	& 6.52    & 4.76 &	-	& -	   & -	   \\

SMC5\_004988	& 207.16376.31	& B2IV  &134         &0.59378 	& 6.14    & 3.42 & 0.59378 	& 6.09    & 3.22 \\
SMC5\_020451	& 207.16374.240	& B2V   &243        &0.86186 	& 21.69   & 5.90 &	-	& -	   & -	   \\
SMC5\_038033	& 207.16147.27	& B2IV  &76        &0.59241 	& 15.14   & 0.61 & 0.59236 	& 10.24   & 0.84 \\
SMC5\_038311	& 207.16376.49	& B2IV  &196        &0.59104 	& 6.85    & 2.67 &	-	& -	   & -	   \\
SMC5\_050662	& 207.16375.56	& B2IV  &161        &1.19411 	& 36.73   & 0.61 & 1.19410 	& 39.20   & 0.67 \\
		&		&       &        & 0.59707 	& 12.79   & 1.06 & 0.59705 	& 15.02   & 1.00 \\
SMC5\_051147	& 207.16318.27	& B2IV  &246        &1.31646 	& 26.10   & 6.08 & 1.31647 	& 26.90   & 6.04 \\
SMC5\_052147	& 207.16204.70	& B2V   &126        &1.04202 	& 17.45   & 3.14 & 1.04196 	& 14.91   & 3.68 \\
SMC5\_087022	& 207.16376.188	& B2V   &145        &0.55267 	& 13.27   & 5.70 &	-	& -	   & -	   \\
\hline
\end{tabular}
\end{table*}

\begin{table*}
\caption{Pulsational characteristics of short-period variable Be stars in the SMC.}
\label{Be-SP}
\centering
\begin{tabular}{ c @{\ \ } c @{\ \ } c @{\ \ } c @ {\ \ } c @{\ \ } c @{\ \ } c @{\ \ } c @{\ \ } c }
\hline
\hline
\multicolumn{2}{ c}{Star iD}&& \multicolumn{3}{c}{B filter}  & \multicolumn{3}{c }{R filter} \\
EIS iD          & MACHO iD	& Sp. T. & Freq	        & Amp	   & Phase & Freq	& Amp	   & Phase \\ 
                &               &        & [\cd]		& [mmag]   & [rad] & [\cd]	& [mmag]   & [rad] \\
\hline
SMC5\_000643	& 207.16318.32	& B3IVe   &3.69095	& 4.76    & 3.83 & 	-	& -	   & -	   \\
		&		&         &1.65525	& 4.63    & 2.80 &1.65555	& 8.86    & 0.68 \\
SMC5\_002232	& 207.16372.22	& B2IIIe  &1.32656	& 25.31   & 6.17 & 1.32661	& 33.51   & 5.59 \\
		&		&         &1.32620	& 14.97   & 1.27 & 1.32616	& 18.03   & 1.88 \\
SMC5\_004201	& 207.16375.39	& B5II-IIIe&1.69005 	& 45.88   & 2.16 & 1.69039 	& 79.76   & 6.05 \\
		&		&         &1.69052 	& 30.58   & 4.24 & 1.69050 	& 55.69   & 2.25 \\
		&		&         &1.69086 	& 24.53   & 4.76 & 1.68997 	& 28.21   & 2.34 \\
SMC5\_011991	& 207.16315.67	& B2IVe    &2.30208 	& 4.38    & 1.01 &	-	& -	   & -	   \\
SMC5\_013978	& 207.16373.58	& B3IIIe   &1.37946 	& 15.50   & 0.48 & 1.37948	& 14.15   & 0.06 \\
		&		&         &0.59335 	& 14.36   & 2.93 & 0.59336	& 15.42   & 2.78 \\
SMC5\_014212	& 207.16259.29	& B2IIIe   &1.27691 	& 23.48   & 4.91 & 1.27693 	& 15.75   & 4.90 \\
SMC5\_014271	& 207.16259.30	& B0IIIe   &4.16878 	& 7.33    & 1.67 &	-	& -	   & -	   \\
		&		&         &4.08090 	& 5.84    & 2.59 &	-	& -	   & -	   \\
		&		&         &4.24210 	& 5.51    & 2.10 &	-	& -	   & -	   \\
SMC5\_014727	& 207.16373.63	& B2IVe   &1.12290 	& 16.65   & 4.04 &	-	& -	   & -	   \\
SMC5\_014878	& 207.16259.41	& B2IIIe  &3.50858 	& 18.93   & 4.12 & 3.50877 	& 8.79    & 2.36 \\
		&		&         &3.50846 	& 11.90   & 2.59 &	-	& -	   & -	   \\
SMC5\_016523	& 207.16316.30	& B2IIIe  &1.29297	& 40.29   & 0.93 & 1.29296 	& 33.43   & 0.95 \\
		&		&         &1.29344	& 23.68   & 3.80 & 1.29341 	& 20.92   & 3.90 \\
SMC5\_016544	& 207.16373.129	& B2IVe   &1.70768 	& 29.54   & 3.27 & 1.70776 	& 32.97   & 2.46 \\
		&		&         &1.65011 	& 22.46   & 1.32 & 1.64988	& 24.58   & 3.08 \\
		&		&         &1.64959 	& 11.36   & 6.23 & 1.65029 	& 12.90   & 0.41 \\
SMC5\_021152	& 207.16147.14	& B2IIIe   &0.98514 	& 19.25   & 2.55 & 0.98513 	& 17.08   & 2.67 \\
		&		&         &1.00442 	& 19.06   & 4.45 & 1.00439 	& 17.01   & 4.67 \\
SMC5\_025718	& 207.16262.86	& B2IIIe  &0.85851 	& 16.26   & 3.86 & 0.85847 	& 13.61   & 4.06 \\
SMC5\_025829	& 207.16376.43	& B2III-IVe&0.85031 	& 7.56    & 2.25 & 0.85019	& 13.93   & 2.46 \\
SMC5\_026689	& 207.16205.141	& B2Ve     &3.82424 	& 10.24   & 5.79 &	-	& -	   & -	   \\
SMC5\_037013	& 207.16315.26	& B2IIIe   &1.18149 	& 15.70   & 2.17 & 1.18154 	& 20.06   & 1.73 \\
		&		&          &1.21712 	& 10.64   & 0.18 & 1.21707 	& 9.48    & 0.14 \\
		&		&          &1.18116 	& 8.39    & 0.01 & 1.18113	& 5.75    & 0.09 \\
SMC5\_037137	& 207.16373.30	&  B2IIIe   &-	& -	   & -	   & 0.38864 	& 6.29    & 3.44 \\
SMC5\_037158	& 207.16316.99	& B1IVe     &3.79973 	& 16.21   & 1.90 &	-	& -	   & -	   \\
		&		&           &3.80111 	& 12.79   & 2.51 &	-	& -	   & -	   \\
		&		&           &3.80020 	& 12.34   & 2.92 &	-	& -	   & -	   \\
		&		&           &3.80187 	& 9.79    & 4.37 &	-	& -	   & -	   \\
SMC5\_037162	& 207.16259.57	& B2IIIe    &0.88531 	& 37.31   & 4.92 & 0.88532 	& 39.17   & 4.79 \\
		&		&           &0.90613 	& 19.21   & 5.61 & 0.90622 	& 17.71   & 4.65 \\
SMC5\_043413	& 207.16315.41	& B2IVe     &2.00731 	& 6.74    & 2.21 & 2.00709 	& 6.21    & 2.84 \\
SMC5\_044898	& 207.16373.132	& B2III-IVe &1.19626 	& 15.28   & 0.14 & 1.19627 	& 18.32   & 6.21 \\
SMC5\_045353	& 207.16259.35	& B2IIIe    &0.59169 	& 33.01   & 2.11 & 0.59168 	& 42.01   & 2.26 \\
		&		&           &0.56293 	& 8.65    & 2.46 &	-	& -	   & -	   \\
		&		&           &1.33684 	& 7.26    & 4.94 &	-	& -	   & -	   \\
SMC5\_048047	& 207.16431.1732& B2IVe     &0.60530 	& 5.86    & 5.52 & 0.60544 	& 5.45    & 4.77 \\
		&		&           &1.15074 	& 5.17    & 2.97 & 1.15909 	& 4.86    & 4.01 \\
SMC5\_049996	& 207.16147.29	& B2IIIe    &1.09797 	& 18.75   & 2.68 & 1.09796 	& 20.05   & 2.71 \\
		&		&           &1.00391 	& 16.19   & 0.82 & 1.00394 	& 18.70   & 0.59 \\
SMC5\_064327 	& 207.16373.51	& B3II-IIIe &1.05025 	& 9.19    & 0.54 &	-	& -	   & -	   \\
SMC5\_074402	& 207.16147.28	& B2IIIe    &1.54904 	& 5.30    & 0.88 &	-	& -	   & -	   \\
		&		&           &1.54956 	& 4.96    & 3.75 &	-	& -	   & -	   \\
SMC5\_079021	& 207.16203.94	& B2IV      &3.48747	& 10.28   & 0.53 & 3.48753	& 8.47    & 2.19 \\
		&		&           &3.45715	& 4.19	   & 2.89 &	-	& -	   & -	   \\
SMC5\_080910	& 207.16262.58	& B2IVe     &1.14776 	& 37.35   & 1.87 & 1.14771 	& 38.84   & 1.93 \\
		&		&           &1.14806 	& 11.65   & 4.34 &	-	& -	   & -	   \\
SMC5\_082042	& 207.16375.41	& B3IIIe    &2.48838 	& 19.06   & 4.87 & 2.48833 	& 12.10   & 5.37 \\
		&		&           &1.16624 	& 16.80   & 2.66 & 1.16627 	& 11.60   & 2.54 \\
SMC5\_082819	& 207.16373.24	& B2IIIe    &	-	& -	   & -	   & 0.38869 	& 7.35    & 3.15 \\
SMC5\_082941	& 207.16203.47	& B3IIIe    &0.62490 	& 38.75   & 1.23 & 0.62499 	& 26.85   & 0.66 \\
		&		&           &0.62470 	& 17.16   & 2.73 & 0.62471 	& 25.60   & 2.54 \\
		&		&           &0.15325 	& 11.47   & 5.86 & 0.15325 	& 8.00    & 5.86 \\
SMC5\_086251	& 207.16319.58	& B2IVe     &4.12332 	& 10.33   & 1.49 &	-	& -	   & -	   \\

\hline
\end{tabular}
\end{table*}

\onlfig{2}{
\begin{figure*}
\centering
\includegraphics[width=17cm]{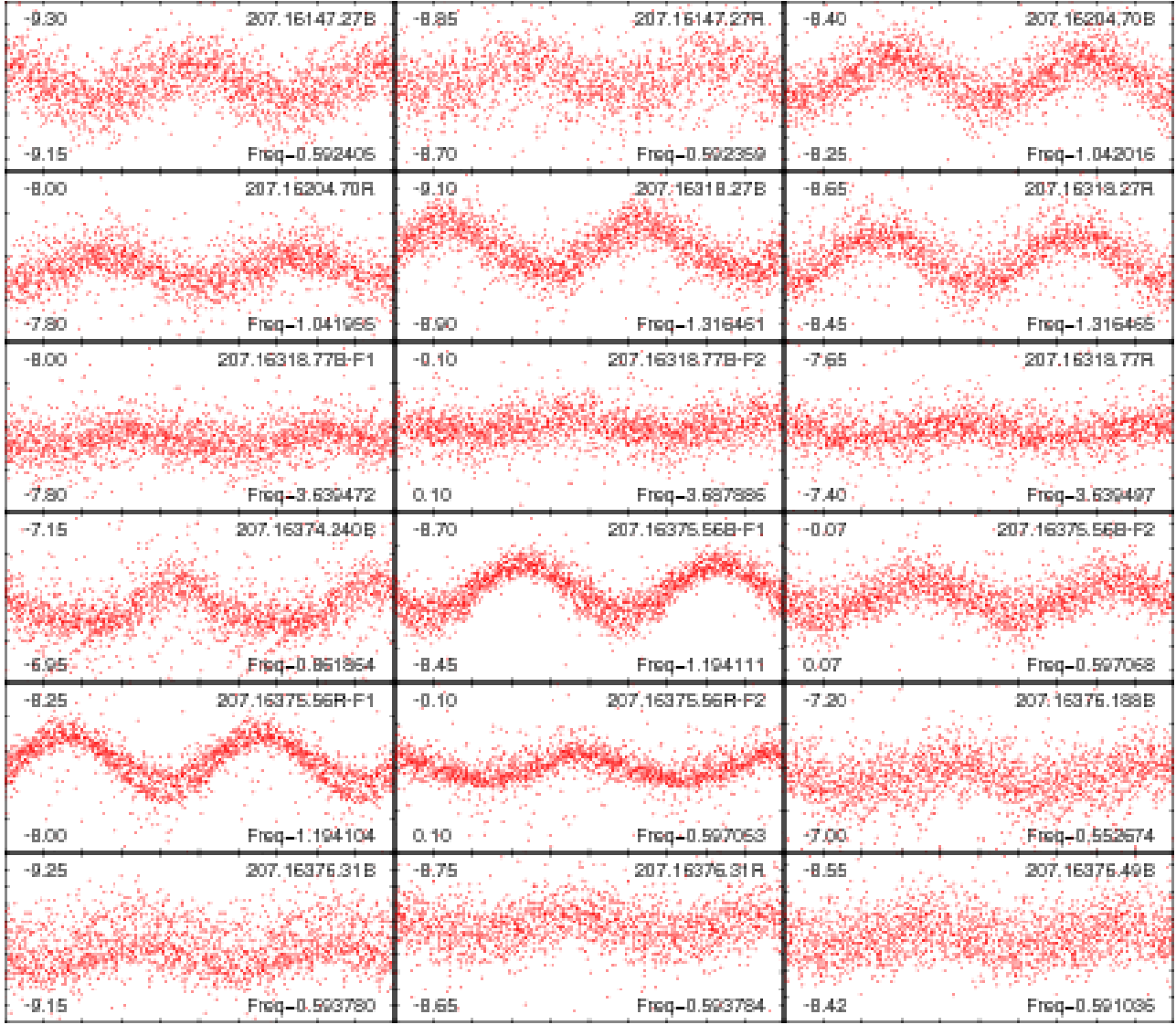}
\caption{Phase plots of absorption-line B stars showing short-term variability folded with the frequency given in each panel. Phases span from 0 to 2. Magnitudes in the B or R filter are denoted inside each panel. When the star is multiperiodic, the frequency used in the plot is given by the label F1, F2, F3 or F4 inside each panel.}
\label{PlotFases-B}
\end{figure*}
}

\onlfig{3}{
\begin{figure*}
\centering
\includegraphics[width=17cm]{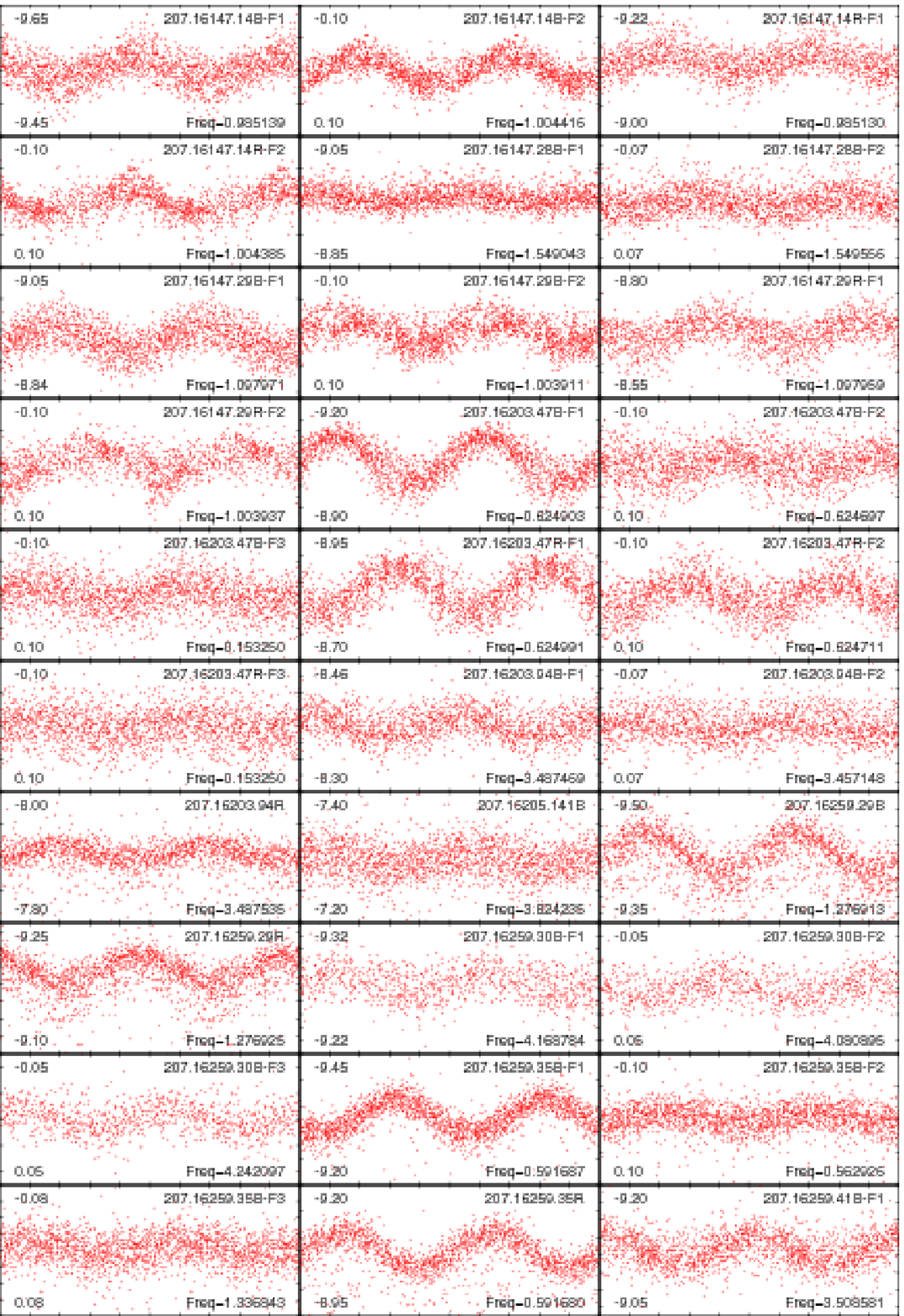}
\caption{Phase plots of Be stars showing short-term variability. Axes and labels as in Fig.~\ref{PlotFases-B}.}
\label{PlotFases-Be-1}
\end{figure*}
}

\onlfig{3}{
\begin{figure*}
\centering
\includegraphics[width=17cm]{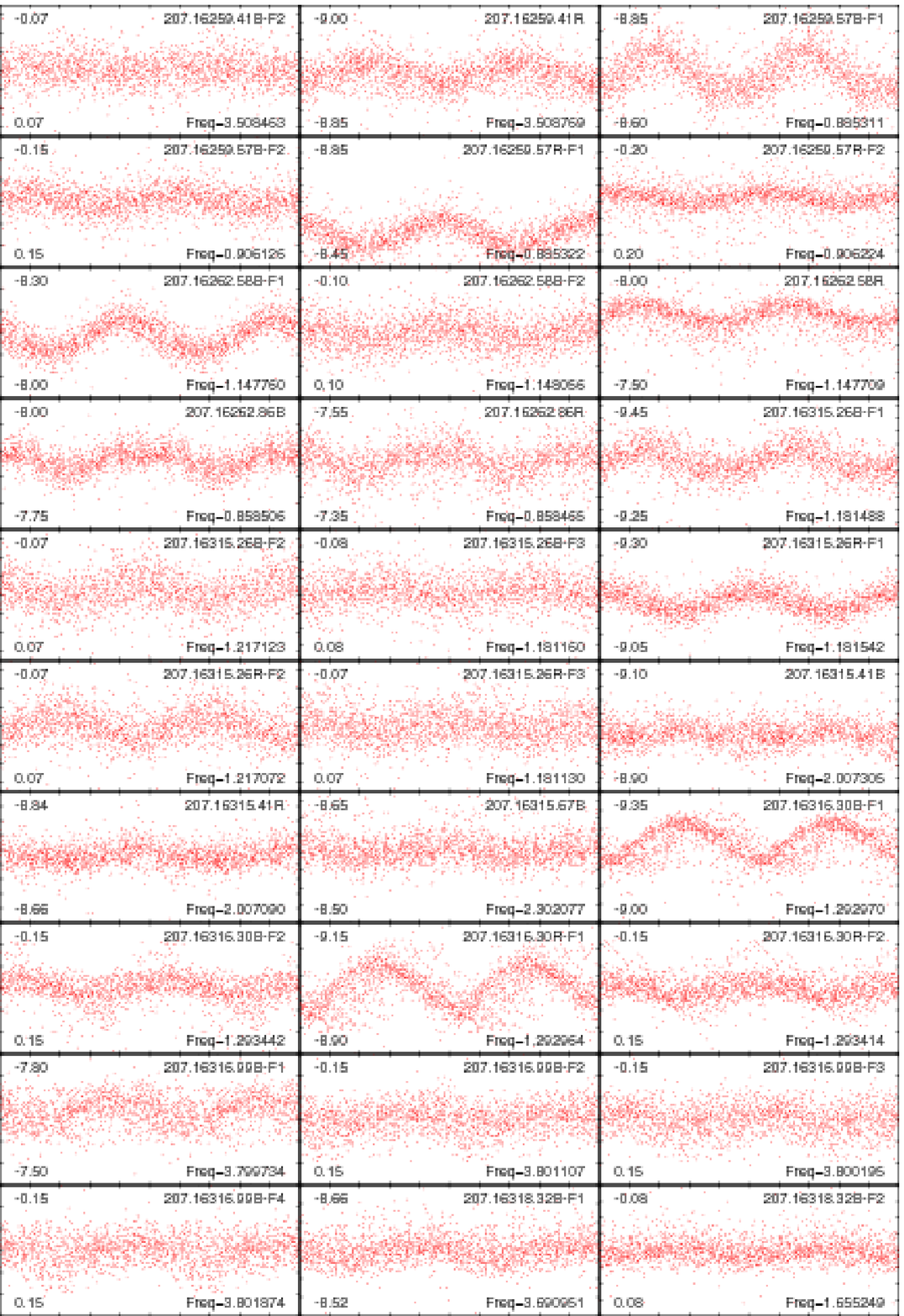}
\caption{Continued.}
\label{PlotFases-Be-2}
\end{figure*}
}

\onlfig{3}{
\begin{figure*}
\centering
\includegraphics[width=17cm]{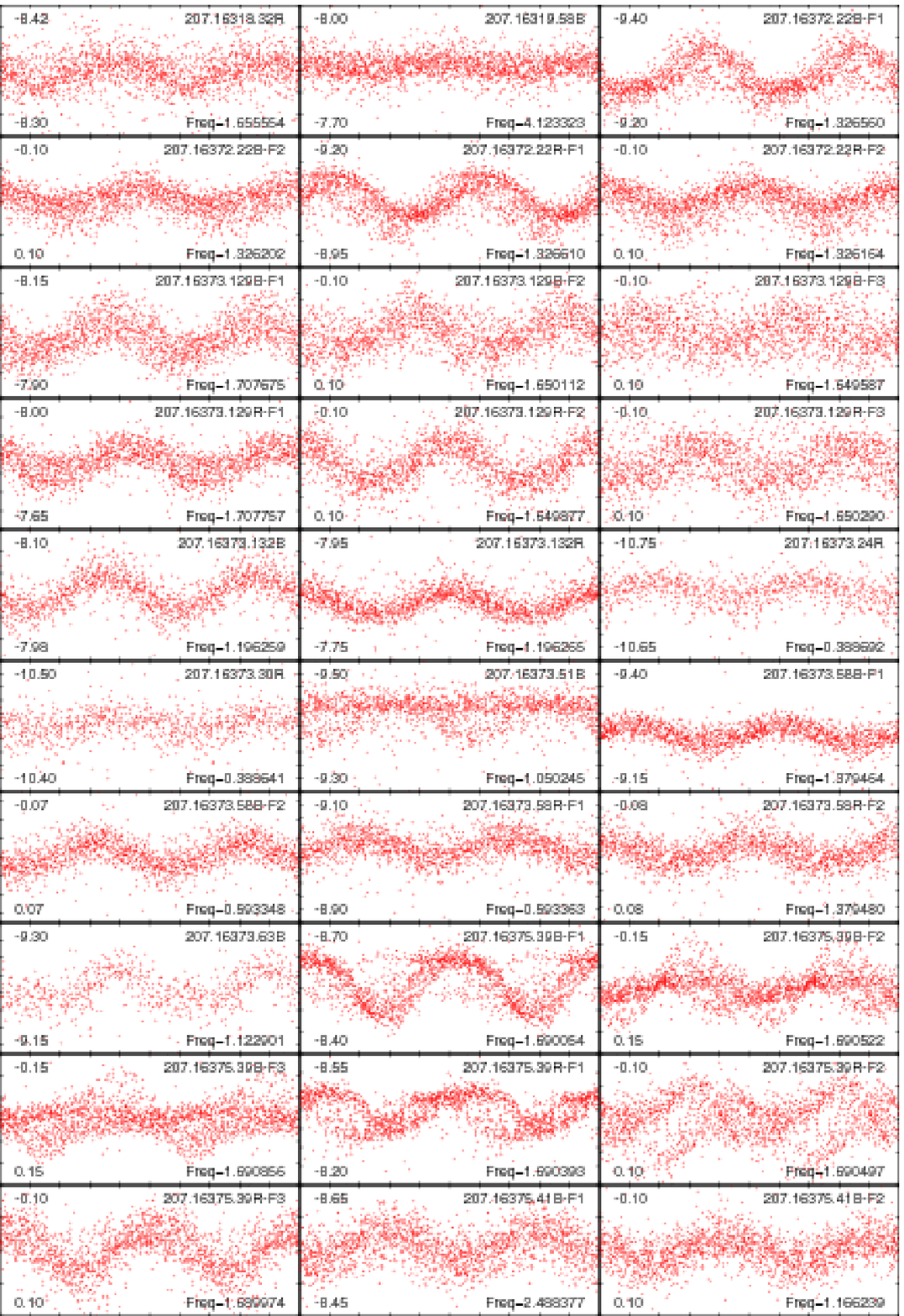}
\caption{Continued.}
\label{PlotFases-Be-3}
\end{figure*}
}

\onlfig{3}{
\begin{figure*}
\centering
\includegraphics[width=17cm]{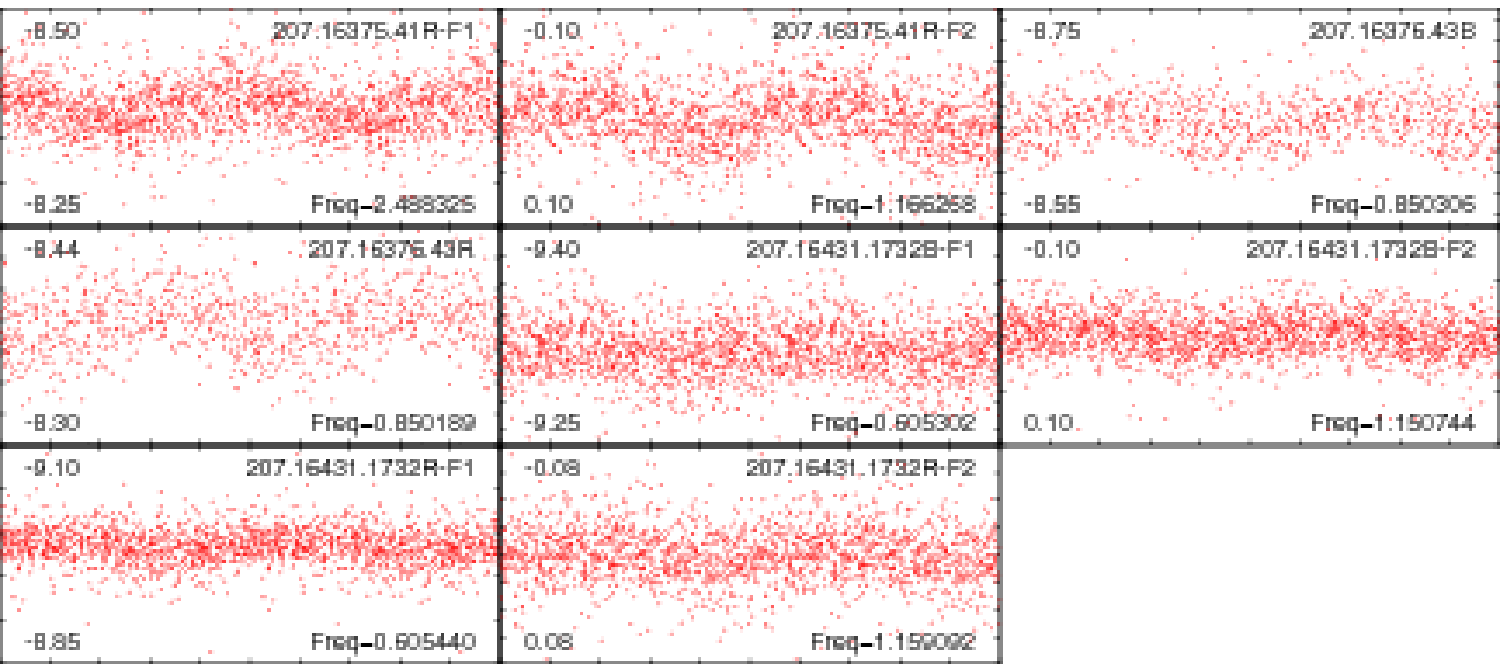}
\caption{Continued.}
\label{PlotFases-Be-4}
\end{figure*}
}

\begin{figure}
\resizebox{\hsize}{!}{\includegraphics[bb=50 50 410 278,clip]{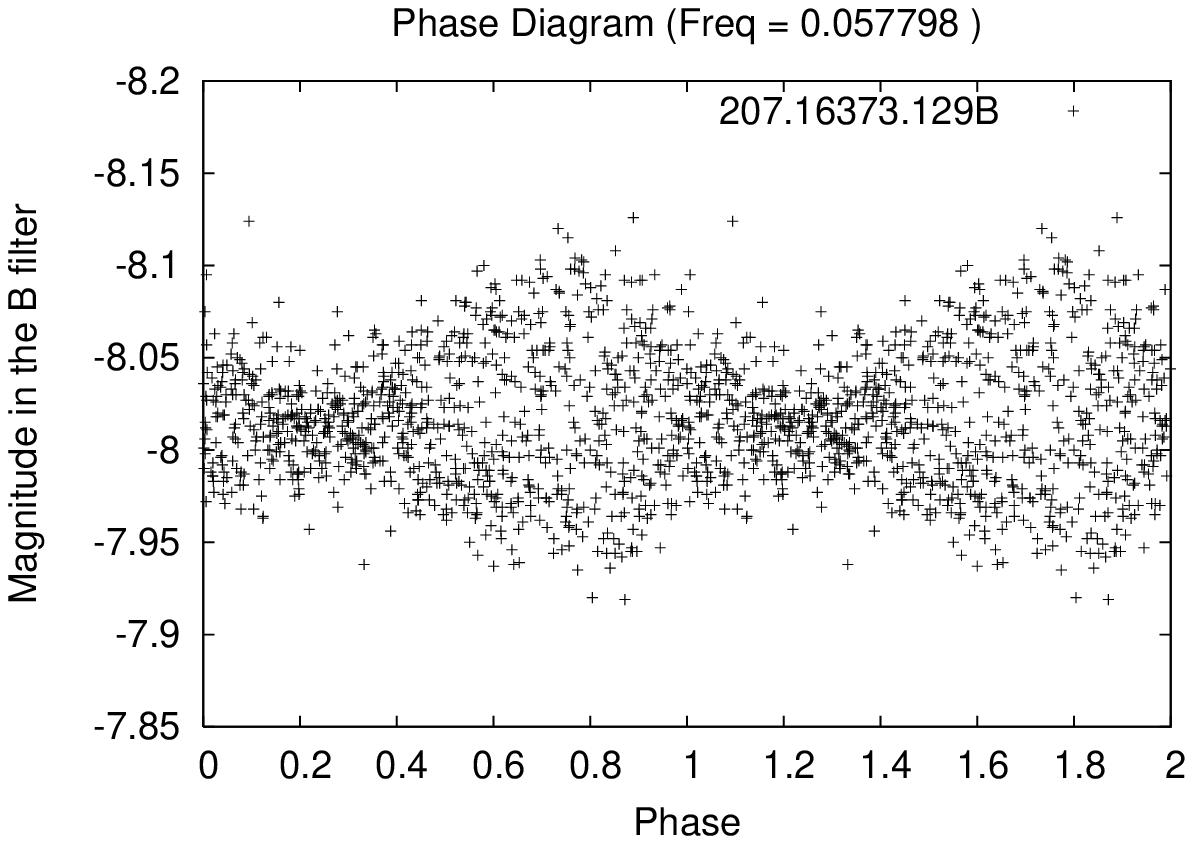}}
\caption{Phase plot of SMC5\_016544 in the B filter folded with the beat frequency (0.05780~\cd). This indicates the presence of close frequencies with similar amplitudes.}
\label{PlotBeating}
\end{figure}

The amplitude variation due to the beating of two close frequencies is clearly seen in the light curves of four multiperiodic variables, when folded with the difference of the two frequencies. This effect occurs when two waves of frequencies $F_{1}$ and $F_{2}$ with similar amplitudes are present in the light curve. The frequency of the resulting wave is the average of the two detected frequencies $\frac{1}{2}(F_{1}+F_{2})$, but the amplitude oscillates with the frequency $F_{1}-F_{2}$. This effect is clearly seen in Fig.~\ref{PlotBeating}, which is an example of a phase diagram folded with the beat frequency ($F_{1}-F_{2}=0.05780$~\cd\ in this case). The plots of the complete set of stars 
showing this pattern and the list of the beat frequencies are presented in Fig.~\ref{PlotBeatingTots}
and Table~\ref{BeatingTable} respectively, available online.
The presence of the beating phenomenon confirms the existence of close frequencies in these stars.

\onltab{4}{
\begin{table*}
\caption{Stars showing the beating phenomenon. The beat frequencies are given in \cd\ in cols.~4 and~5.}
\label{BeatingTable}
\centering
\begin{tabular}{c c c c c}
\hline\hline
\multicolumn{2}{c}{Star iD} & Sp. T. & \multicolumn{2}{c}{Beat frequency }\\
EIS iD	& MACHO iD	& 	& B filter	& R filter\\
\hline
SMC5\_004326	& 207.16318.77	& B	& 0.04841		& -		\\
SMC5\_016544	& 207.16373.129	& Be	& 0.05780		& 0.05790	\\
SMC5\_021152	& 207.16147.14	& Be	& 0.01928		& 0.01926	\\
SMC5\_045353	& 207.16259.35	& Be	& 0.02876		& -		\\
\hline
\end{tabular}
\end{table*}
}

\onlfig{5}{
\begin{figure*}
\centering
\begin{tabular}{cc}
\resizebox{80mm}{!}{\includegraphics{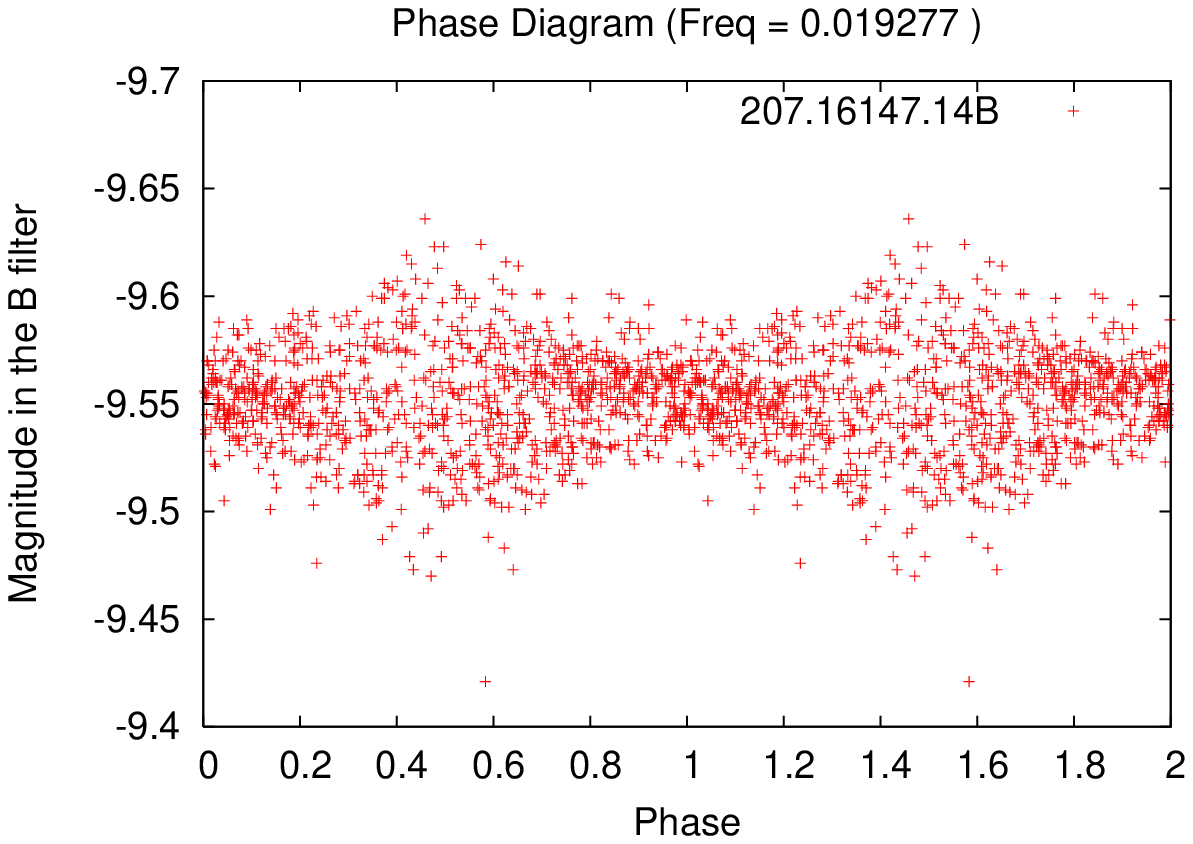}} &
\resizebox{80mm}{!}{\includegraphics{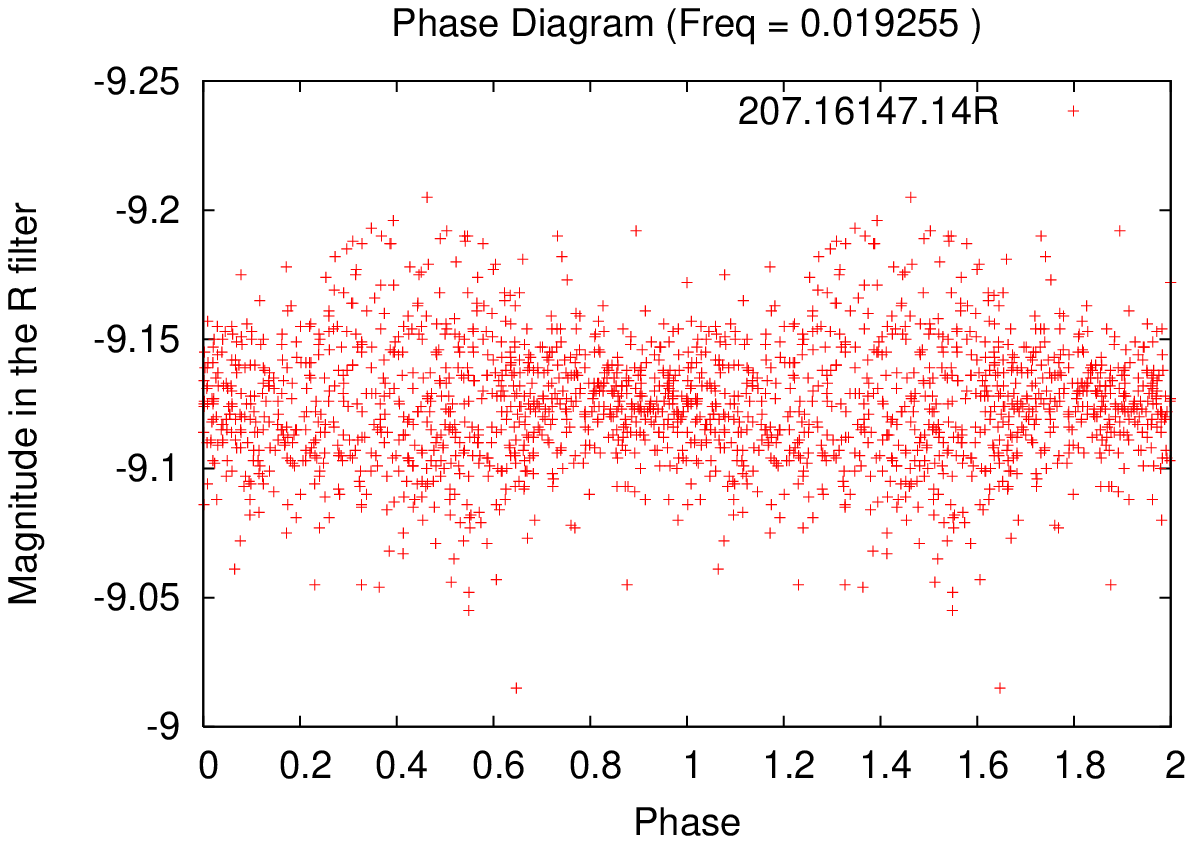}} \\
\resizebox{80mm}{!}{\includegraphics{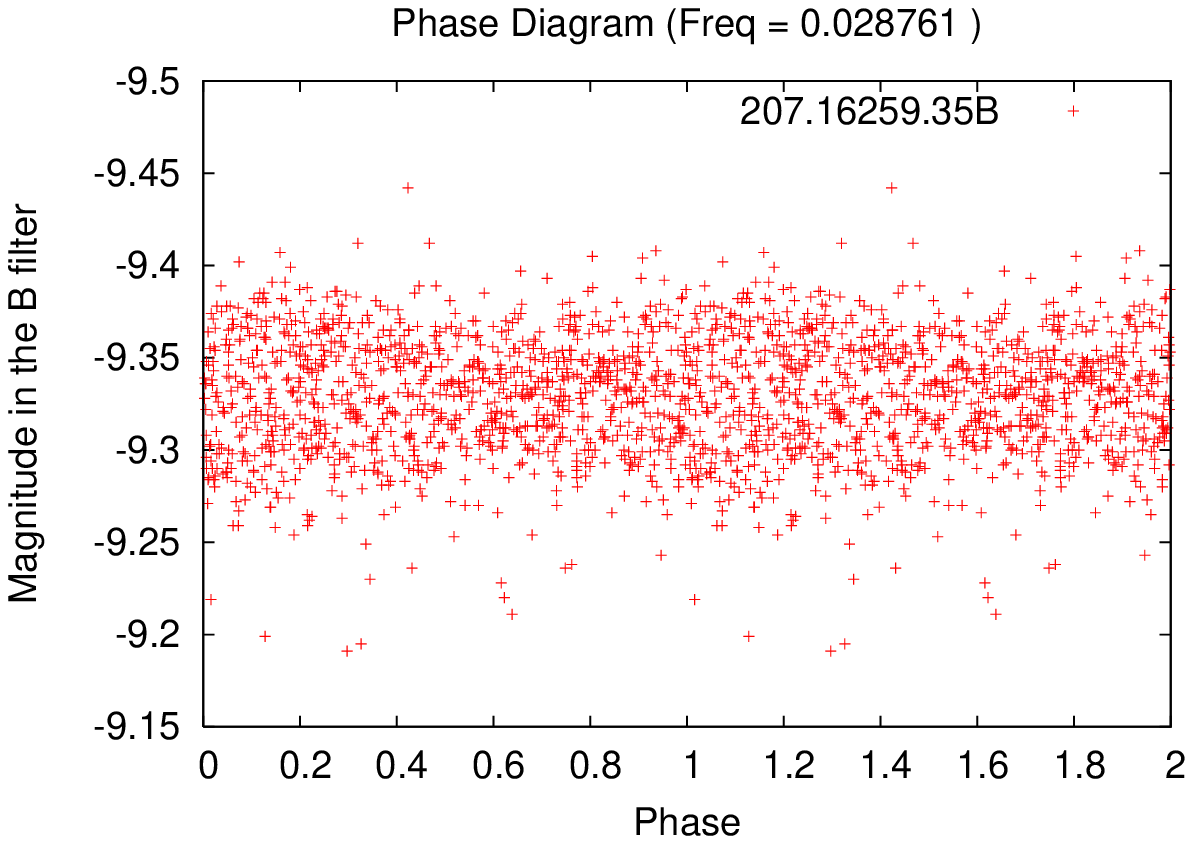}} &
\resizebox{80mm}{!}{\includegraphics{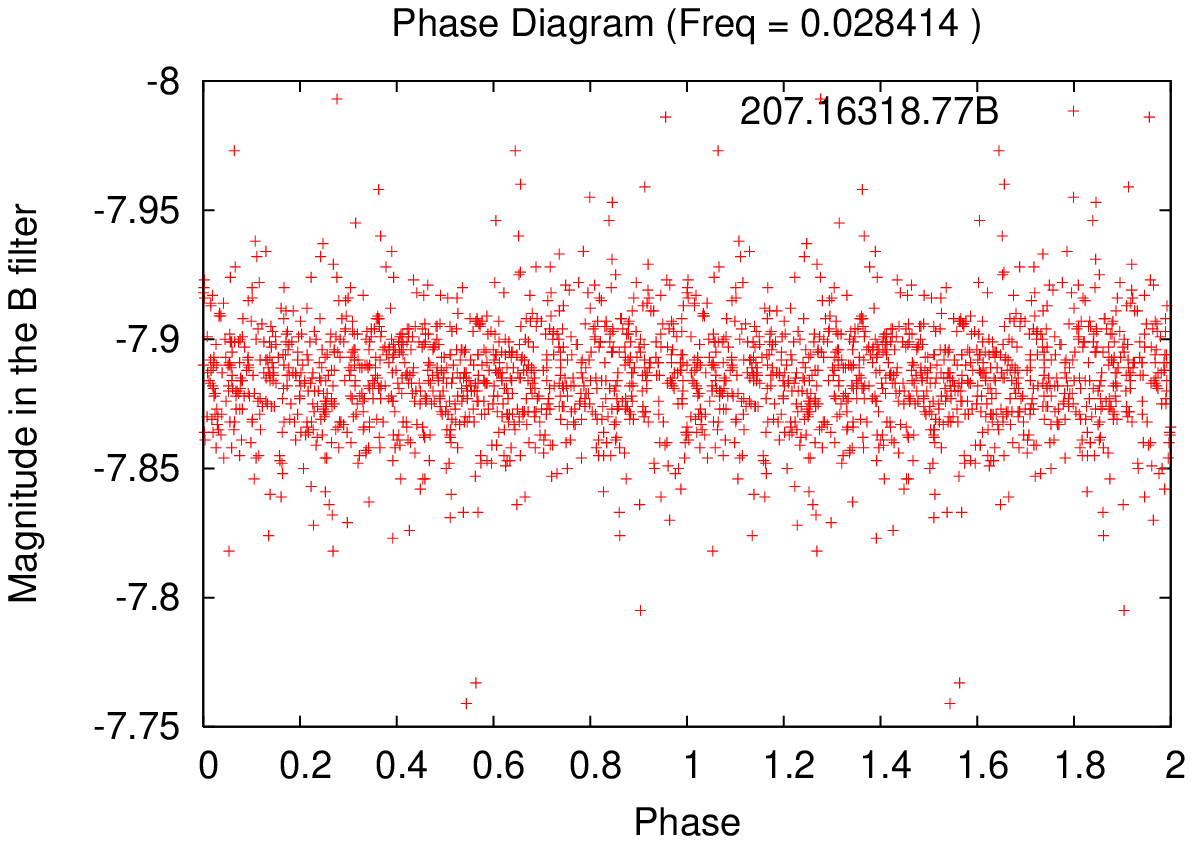}} \\
\resizebox{80mm}{!}{\includegraphics{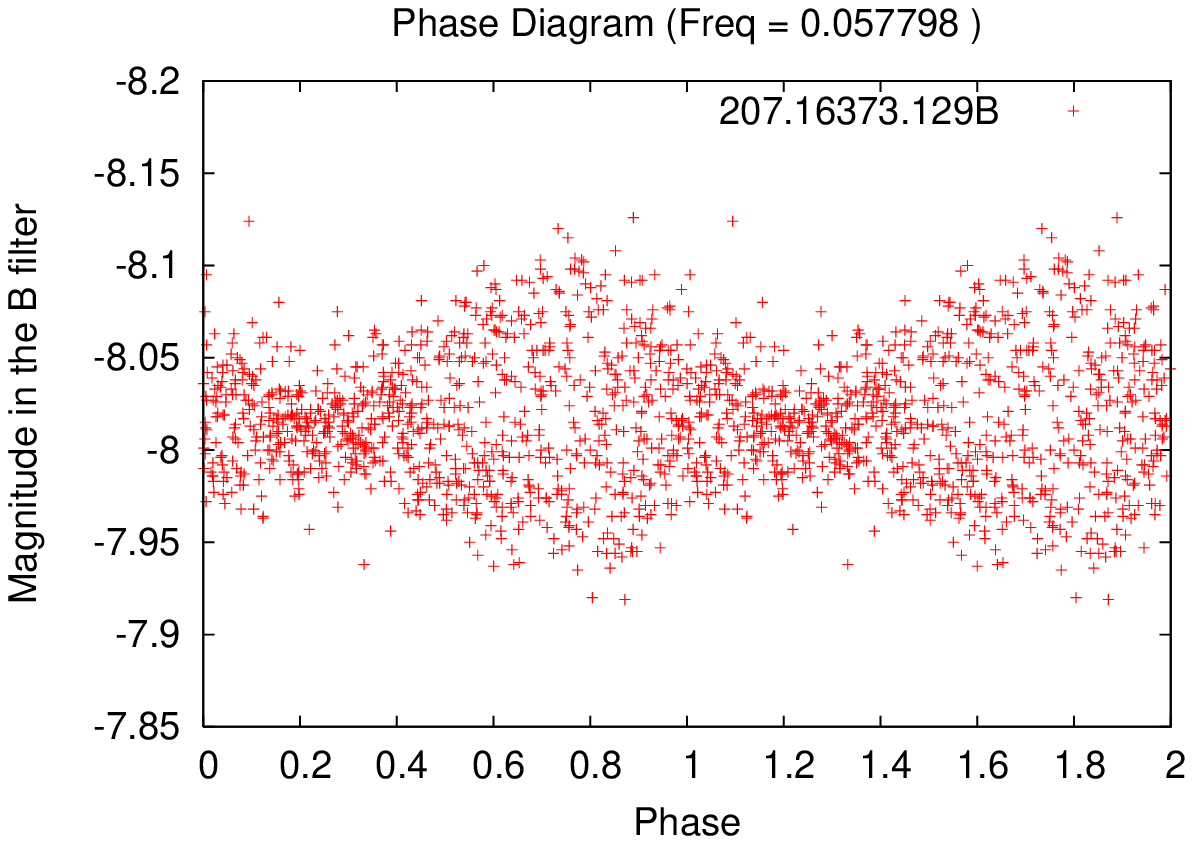}} &
\resizebox{80mm}{!}{\includegraphics{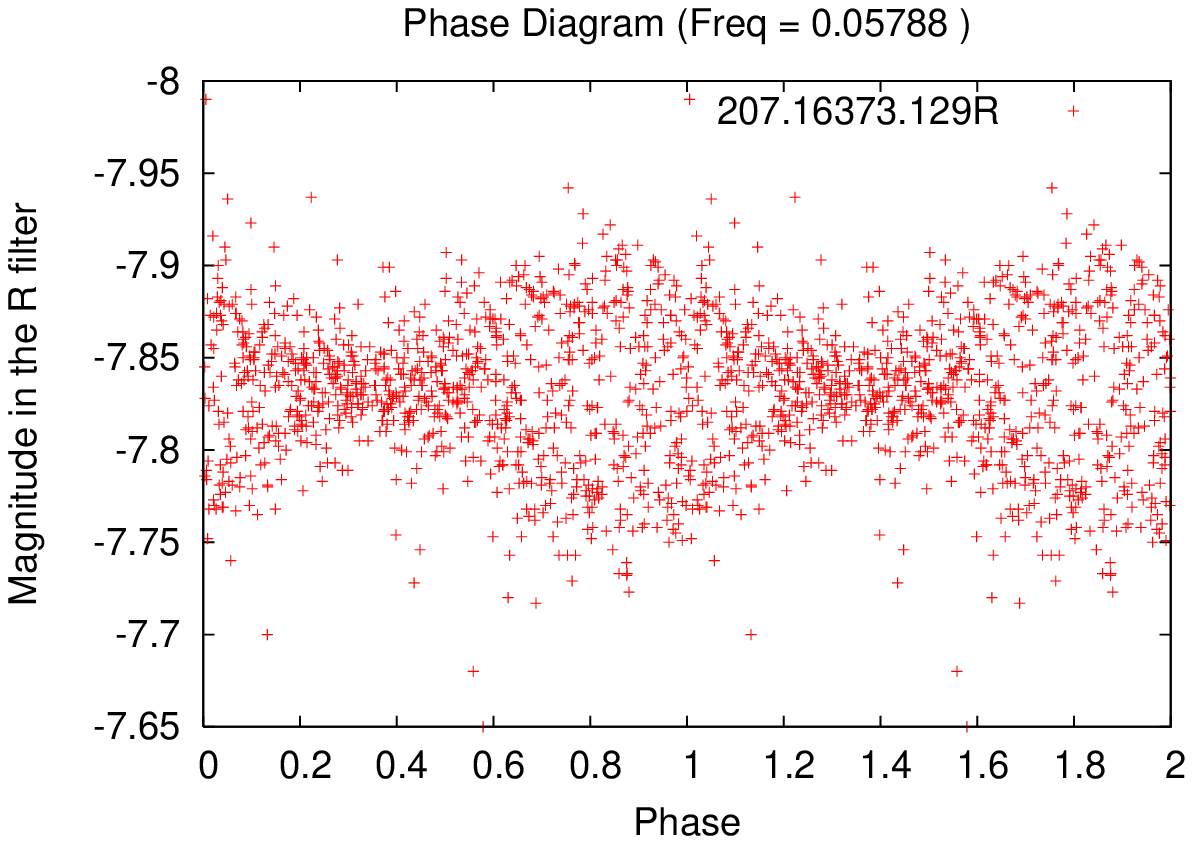}} \\
\end{tabular}
\caption{Phase plots of the stars showing the beating phenomenon.}
\label{PlotBeatingTots}
\end{figure*}
}

\subsection{Long-term variability}

In addition to the short-term variability, we have found periodic or quasi-periodic variations with frequencies lower than 0.333~\cd\ for 6 Be stars. The frequencies, amplitudes and phases are presented in Table~\ref{Be-LP}, and the folded light curves in Fig.~\ref{PlotFases-Be-LP}, both available online.

One of these stars, SMC5\_016161, has been discussed by \citet{Martayan2007}, who showed that it is an spectroscopic binary, and interpreted the light curve as produced by ellipsoidal variability.

\onltab{5}{
\begin{table*}
\caption{Long-period variable Be stars in the SMC.}
\label{Be-LP}
\centering
\begin{tabular}{ c @{\ \ } c @{\ \ } c @ {\ \ } c @{\ \ } c @{\ \ } c @{\ \ } c @{\ \ } c }
\hline
\hline
\multicolumn{2}{ c}{Star iD}& \multicolumn{3}{c}{B filter}  & \multicolumn{3}{c }{R filter} \\
EIS iD & MACHO iD	& Period	& Amp	   & Phase & Period	& Amp	   & Phase \\ 
       &	        & [d]		& [mmag]   & [rad] & [d]	& [mmag]   & [rad] \\
\hline
SMC5\_004509	& 207.16375.57	& 4.67  	& 13.08   & 1.08 &	-	& -	   & -	   \\
SMC5\_016461	& 207.16316.21	& 27.16	        & 88.80	  & 5.84 & 27.15      & 85.46   & 5.57   \\
SMC5\_041410	& 207.16258.73	& 961.54 	& 16.28   & 5.99 & 823.05 	& 36.79   & 3.86 \\
SMC5\_052688	& 207.16319.11	& 8.13 	& 5.06    & 6.02 &	-	& -	   & -	   \\
SMC5\_078440	& 207.16376.22	& 1162.79	& 117.32  & 1.16 & 1129.94	& 210.84   & 0.84 \\
SMC5\_082379	& 207.16204.182	& 9.73 	& 10.96   & 2.54 & 9.73 	& 16.12   & 2.81 \\

\hline
\end{tabular}
\end{table*}
}

\onlfig{6}{
\begin{figure*}
\centering
\includegraphics[width=17cm]{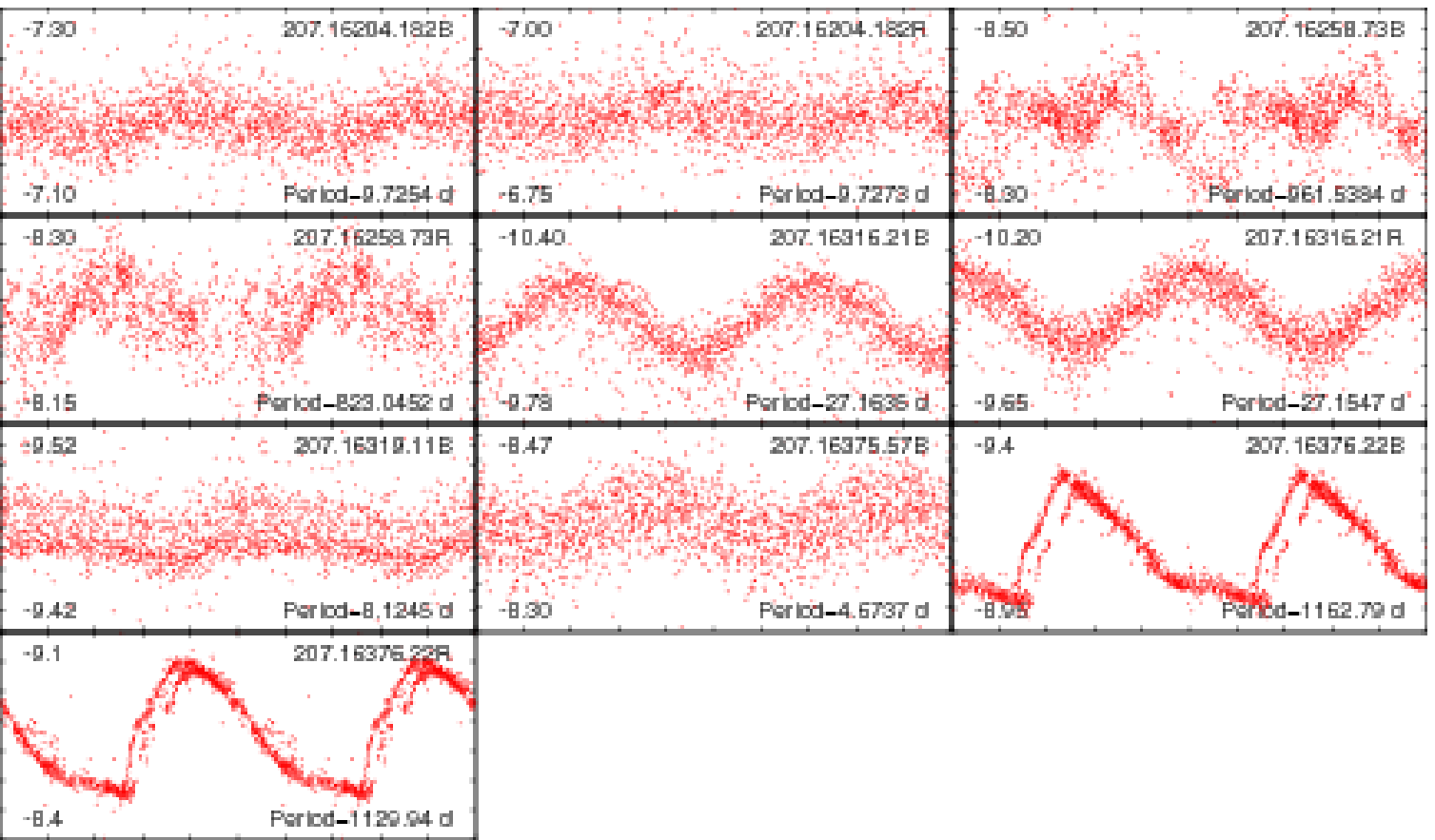}
\caption{Phase plots of the Be stars showing long-term periodic variability. Axes and labels as in Fig.~\ref{PlotFases-B}.}
\label{PlotFases-Be-LP}
\end{figure*}
}

\subsection{Eclipsing binaries}

The light curves of 3 absorption-line B and 1 Be stars are characteristic of eclipsing binaries.  
Periods for each binary star are listed in Table~\ref{BinTab} and phase diagrams are depicted in Figs.~\ref{PlotFases-Binarias-B} and~\ref{PlotFases-Binarias-Be} for B and Be stars respectively (only available online).

\onltab{6}{
\begin{table*}
\caption{Eclipsing binaries in the SMC. Periods are given in days.}
\label{BinTab}
\centering
\begin{tabular}{cccc}
\hline
\hline
EIS iD	& MACHO iD	& Sp. T. & Period  	\\
\hline
SMC5\_003809	& 207.16203.296	& B& 0.8817			\\
SMC5\_004534	& 207.16318.41	& B& 4.0513			\\
SMC5\_015429	& 207.16316.160	& Be& 0.6523			\\
SMC5\_053563	& 207.16205.114	& B& 1.0148			\\
\hline
\end{tabular}
\end{table*}
}

\onlfig{7}{
\begin{figure*}
\centering
\includegraphics[width=17cm]{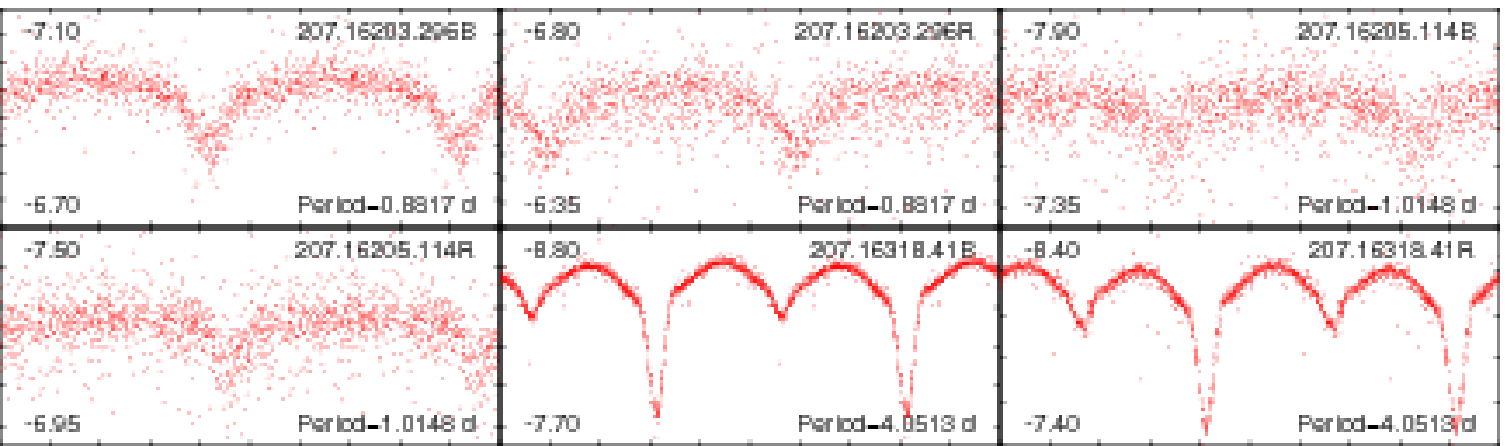}
\caption{Phase plots of eclipsing binaries in the sample of B stars. Axes and labels as in Fig.~\ref{PlotFases-B}.}
\label{PlotFases-Binarias-B}
\end{figure*}
}

\onlfig{8}{
\begin{figure*}
\hspace{0.6cm}
\includegraphics[width=11.5cm]{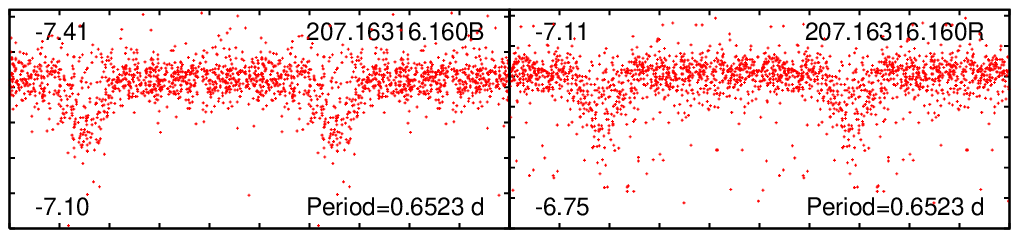}
\caption{The same as Fig.~\ref{PlotFases-Binarias-B} for an eclipsing Be star.}
\label{PlotFases-Binarias-Be}
\end{figure*}
}

\subsection{Irregular variability}

We have found a total of 5 Be stars showing outbursts or irregular variations (see Table~\ref{OutTab}). In 4 of them these variations prevented us from performing a frequency analysis to search for short-term variability, as stated before. 
In Fig.~\ref{OutCompo} (available online) we show the irregular light curves of these Be stars.

The star SMC5\_079021 presents a photometric outburst at the end of the MACHO light curve. 
The data before the outburst does not show significant irregular variability, and therefore we have been 
able to perform a frequency analysis using only the first part of the light curve. Short-period
multiperiodic variability has been detected (see Table~\ref{Be-SP}).

\onltab{7}{
\begin{table*}
\caption{Be stars showing irregular variability or outbursts.}
\label{OutTab}
\centering
\begin{tabular}{cc}
\hline
\hline
EIS iD 		& MACHO iD \\
\hline
SMC5\_073581	& 207.16373.88 \\
SMC5\_074305	& 207.16317.116 \\
SMC5\_079021	& 207.16203.94 \\
SMC5\_190576	& 207.16372.13 \\
SMC5\_744471	& 207.16204.11 \\
\hline
\end{tabular}
\end{table*}
}

\onlfig{9}{
\begin{figure*}
\centering
\begin{tabular}{cc}
\resizebox{90mm}{!}{\includegraphics{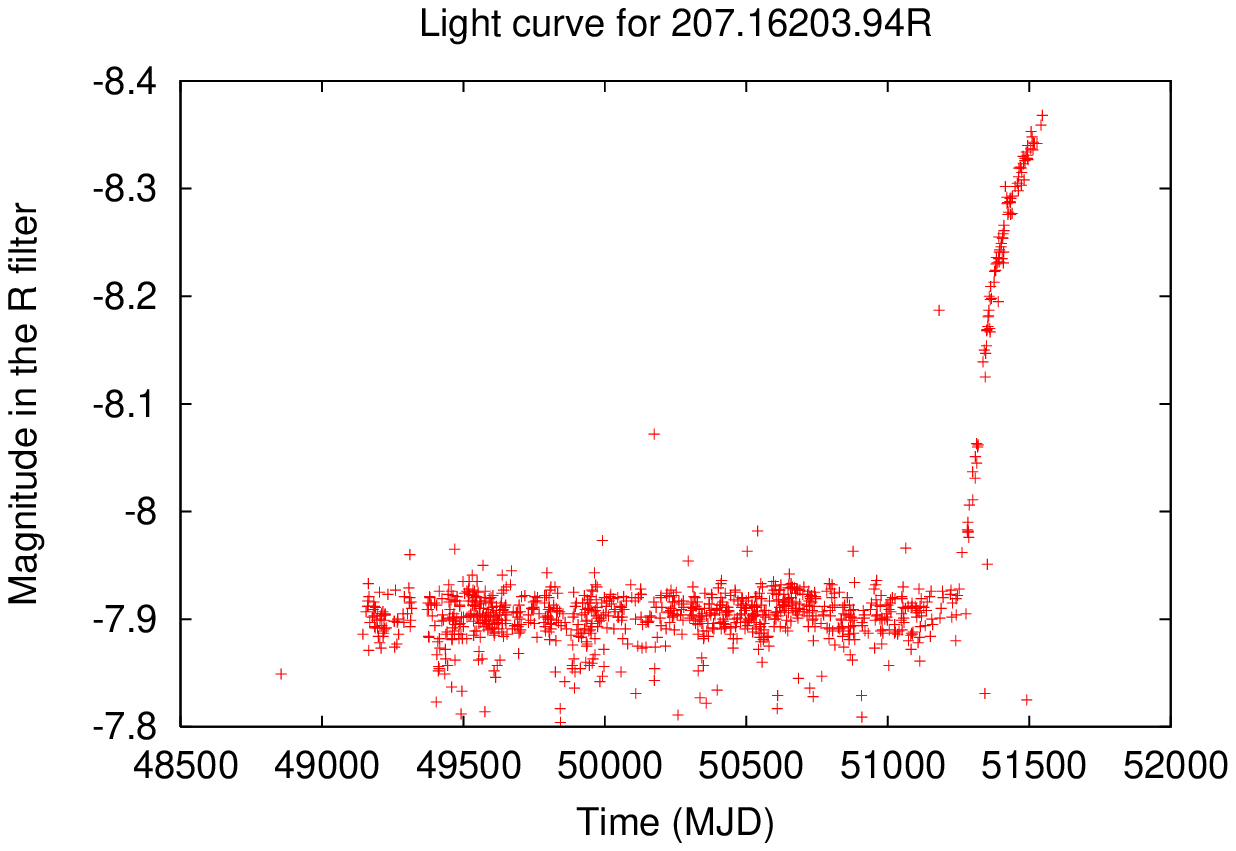}} &
\resizebox{90mm}{!}{\includegraphics{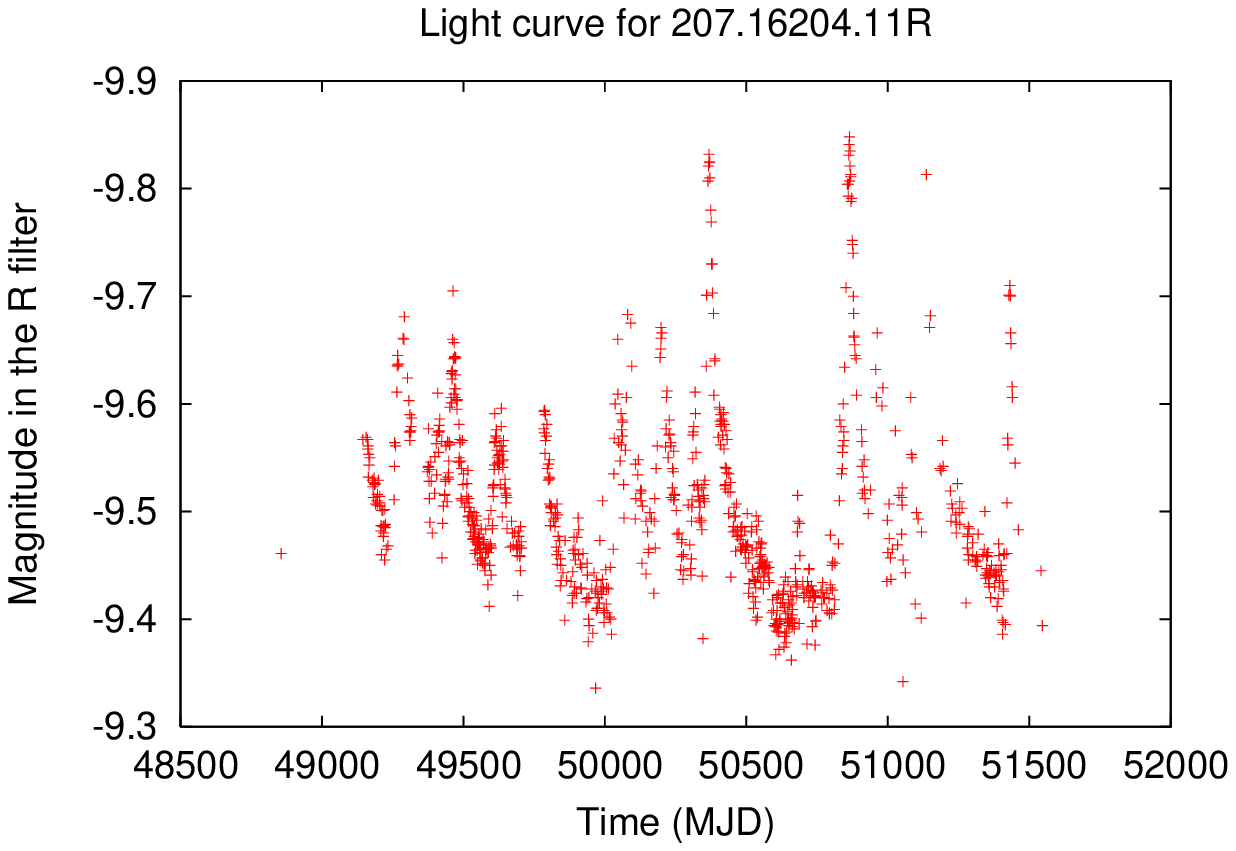}} \\
\resizebox{90mm}{!}{\includegraphics{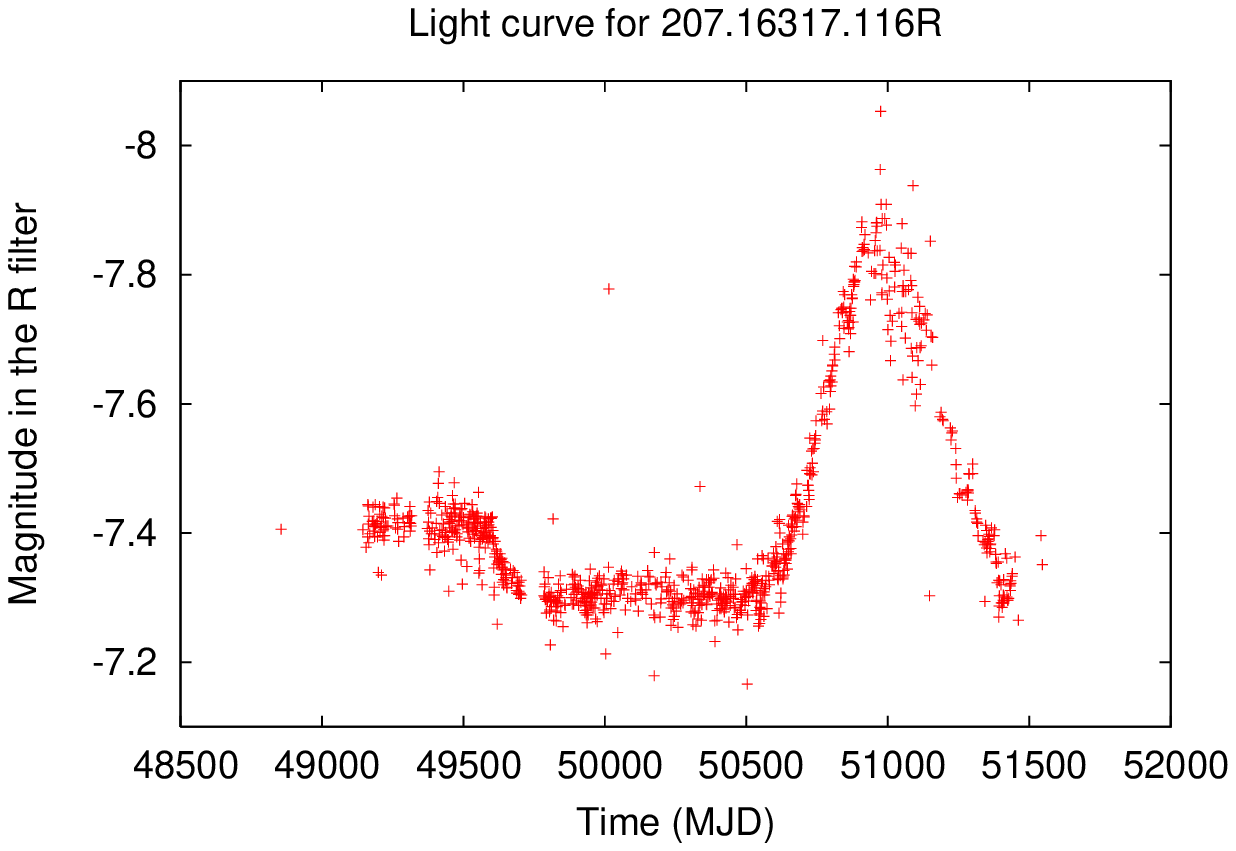}} &
\resizebox{90mm}{!}{\includegraphics{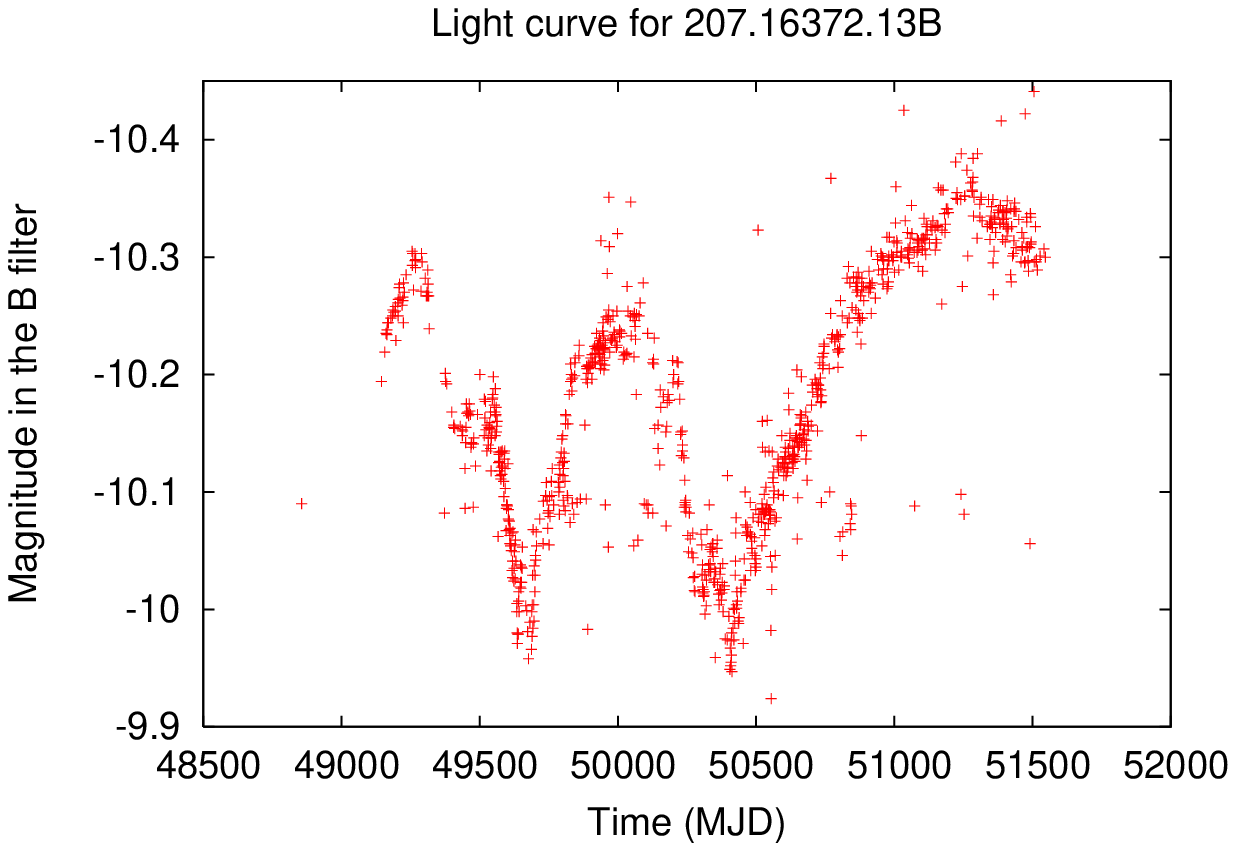}} \\
\multicolumn{2}{c}{
\resizebox{90mm}{!}{\includegraphics{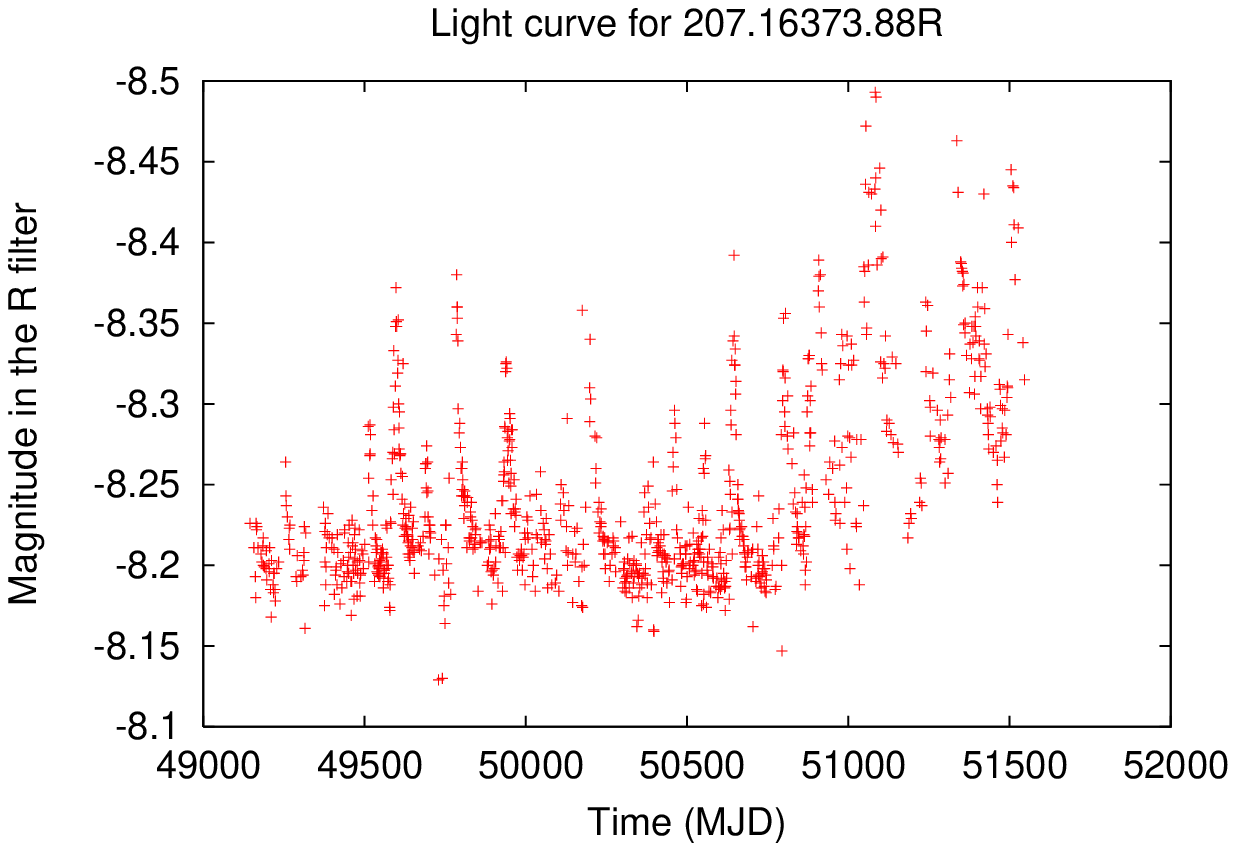}} }\\
\end{tabular}
\caption{Light curves of the Be stars showing outbursts and/or irregular variability.}
\label{OutCompo}
\end{figure*}
}

\section{Discussion}

\subsection{Pulsating B stars}

\begin{figure}
\resizebox{\hsize}{!}{\includegraphics[bb=50 50 410 284,clip]{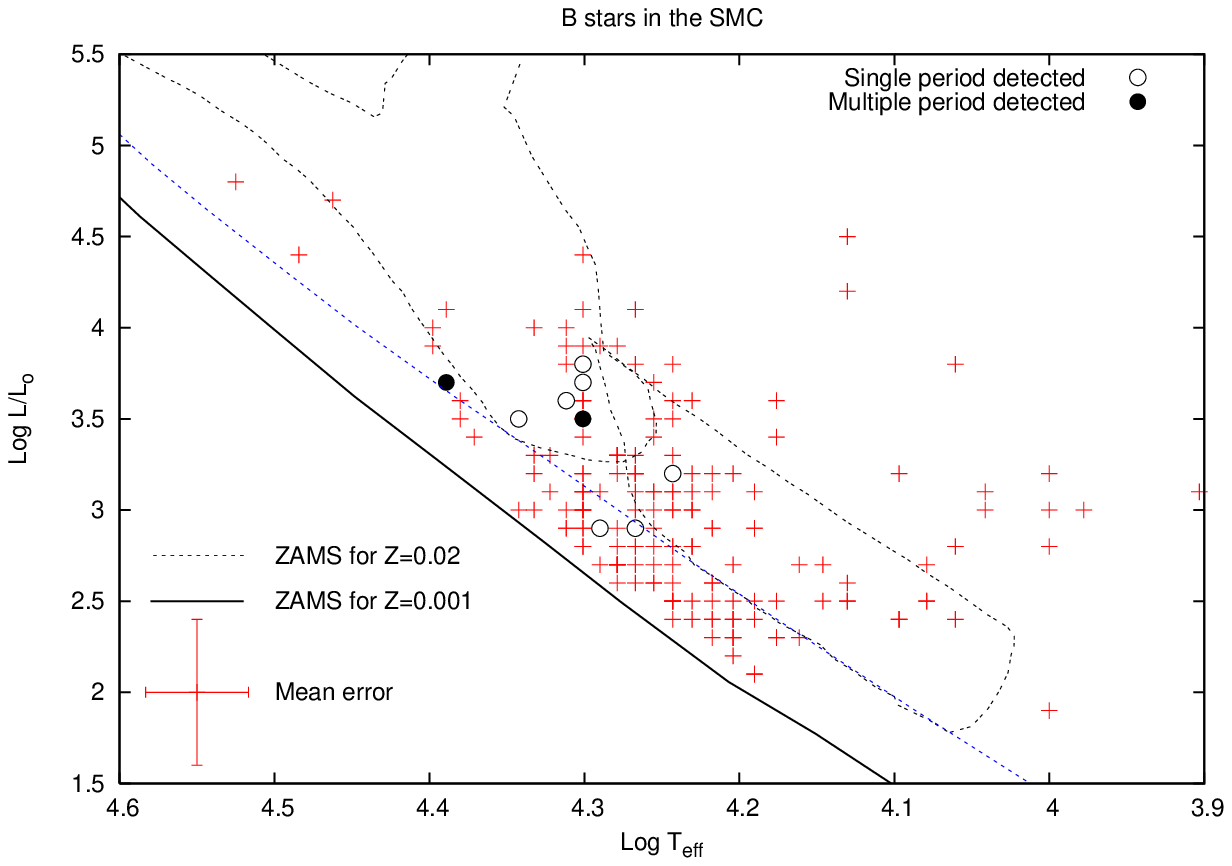}}
\caption{Location of the B star sample in the theoretical HR diagram: single crosses represent stars in our sample, the empty circles represent single period detection and filled ones multiple period detection. The $\beta$ Cephei and the SPB boundaries (dashed lines) at solar metallicity have been taken from \citet{1999AcA....49..119P}. The ZAMS at solar and SMC metallicity 
are from \citet{1992A&AS...96..269S}.}
\label{H-RB}
\end{figure}

We have found 9 short-period variable absorption-line B stars. Their pulsational characteristics are presented in Table~\ref{B-SP}. In Fig.~\ref{H-RB} we have represented their positions in the theoretical HR diagram. The values of $\log T_{\mathrm{eff}}$ and $\log L/L_{\odot}$, as well as the spectral type and $V\!\sin i$, have been taken from M07. For comparison, we have also represented the theoretical boundaries of the $\beta$ Cephei and SPB instability strips and the ZAMS line for solar metallicity. All pulsating B stars are restricted to a narrow range of temperatures, between $\log T_{\mathrm{eff}}=4.24$ and $4.39$~K.

\begin{figure}
\resizebox{\hsize}{!}{\includegraphics{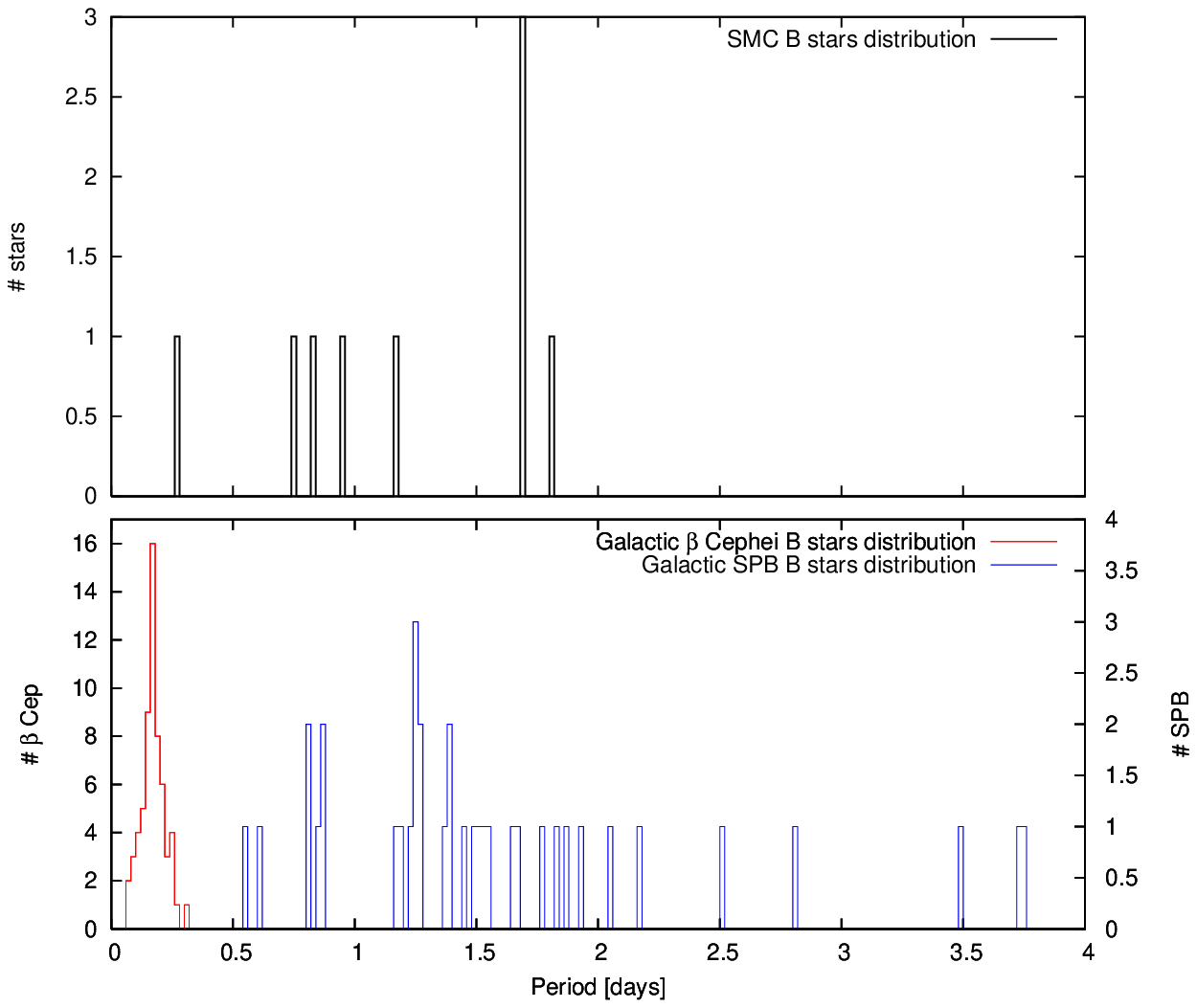}}
\caption{Period distribution of the pulsating B stars found in our SMC sample (top panel) compared to the galactic $\beta$ Cephei and SPB period distributions (bottom panel). All stars in our sample but one have periods longer than 0.5 days, characteristic of SPB stars.}
\label{Histo-B}
\end{figure}

\begin{figure}
\centering
\resizebox{0.9\hsize}{!}{\includegraphics{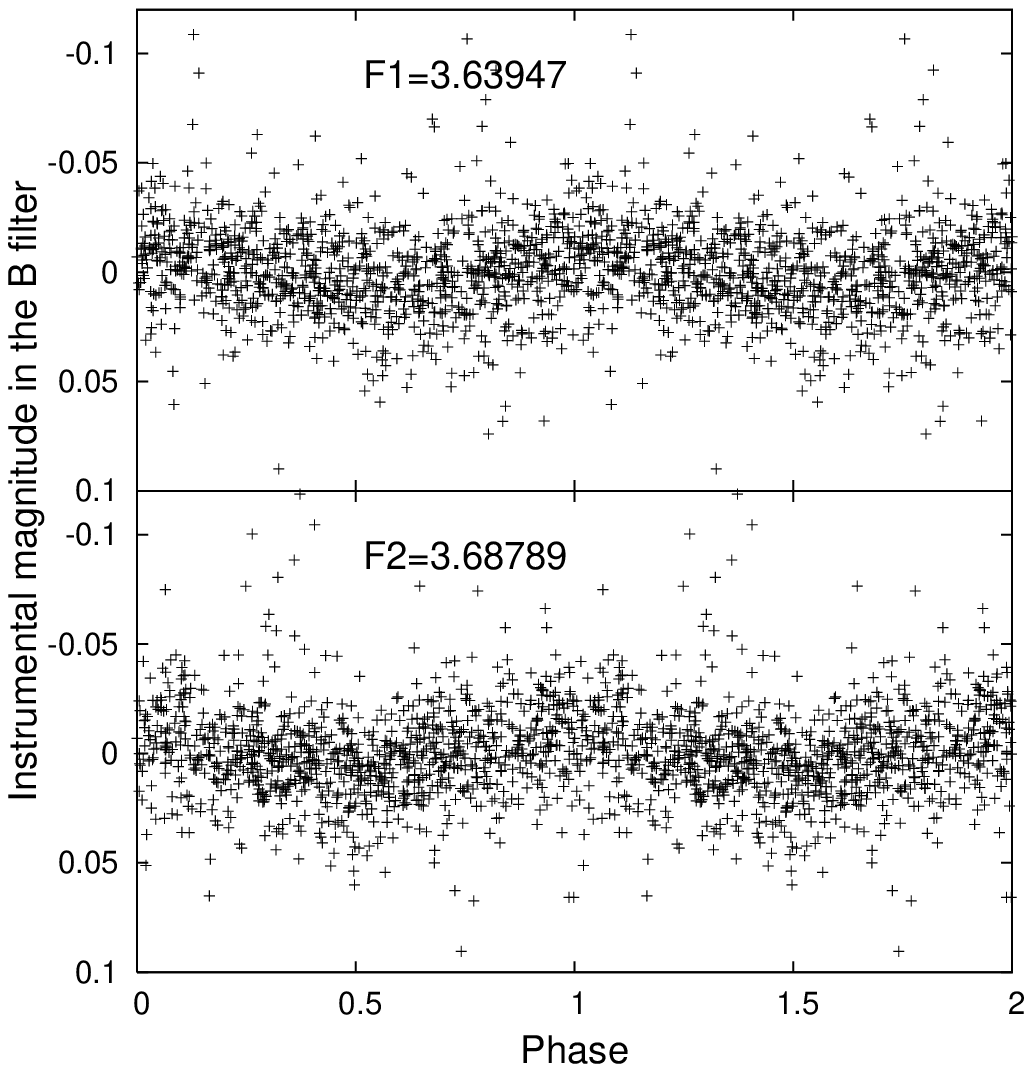}}
\caption{Phase diagram of the $\beta$ Cephei star SMC5\_004326 folded with the two detected frequencies.
}
\label{fig:phase:beta}
\end{figure}

In the bottom panel of Fig.~\ref{Histo-B} we display the period distribution of $\beta$ Cephei and SPB stars in the Galaxy, from \citet{2005ApJS..158..193S} and \citet{2002ASPC..259..196D} respectively.
In the top panel we have represented the period distribution of our sample of pulsating B stars in the SMC. All stars but one have periods longer than 0.5 days, characteristic of SPB stars. Star SMC5\_004326 has two close periods in the range of the p-mode pulsators in the Galaxy. In addition, it is the hottest star in our sample ($\log T_{\mathrm{eff}}=4.39$~K). 
We propose that this star is a candidate $\beta$ Cephei variable in the SMC.
In Fig.~\ref{fig:phase:beta} we display the phase diagrams folded with the two detected frequencies.

However, the periods found are somewhat too long to
correspond to p-modes of a little evolved star. This may well be a high-order
g-mode pulsator, ie. a SPB star. If this is the case, the SPB instability strip
in the SMC is considerably shifted towards higher temperatures than predicted by the 
current models, as discussed in more detail below.

In addition, SMC5\_004326 has a large rotational velocity $V\!\sin i=370\pm30$ km~s$^{-1}$, while most $\beta$ Cephei stars in the Galaxy are known to be slow rotators \citep{2005ApJS..158..193S}. However, these authors 
include in their list several fast-rotating bona fide $\beta$ Cephei stars. They also suggest that the fact that most $\beta$ Cephei stars are slow rotators can be a selection effect, as the highest-amplitude pulsators have the lowest rotational velocities.

If SMC5\_004326 is indeed a $\beta$ Cephei star, this would constitute an unexpected result, as the current stellar models do not predict p-mode pulsations at the SMC metallicity. 
However, as the assumed SMC metallicity is an average value, this particular case 
might correspond to a star with a higher metal content. A spectroscopic 
determination of its metallicity would help resolve this issue.

\begin{figure}
\resizebox{\hsize}{!}{\includegraphics[bb=50 50 410 284,clip]{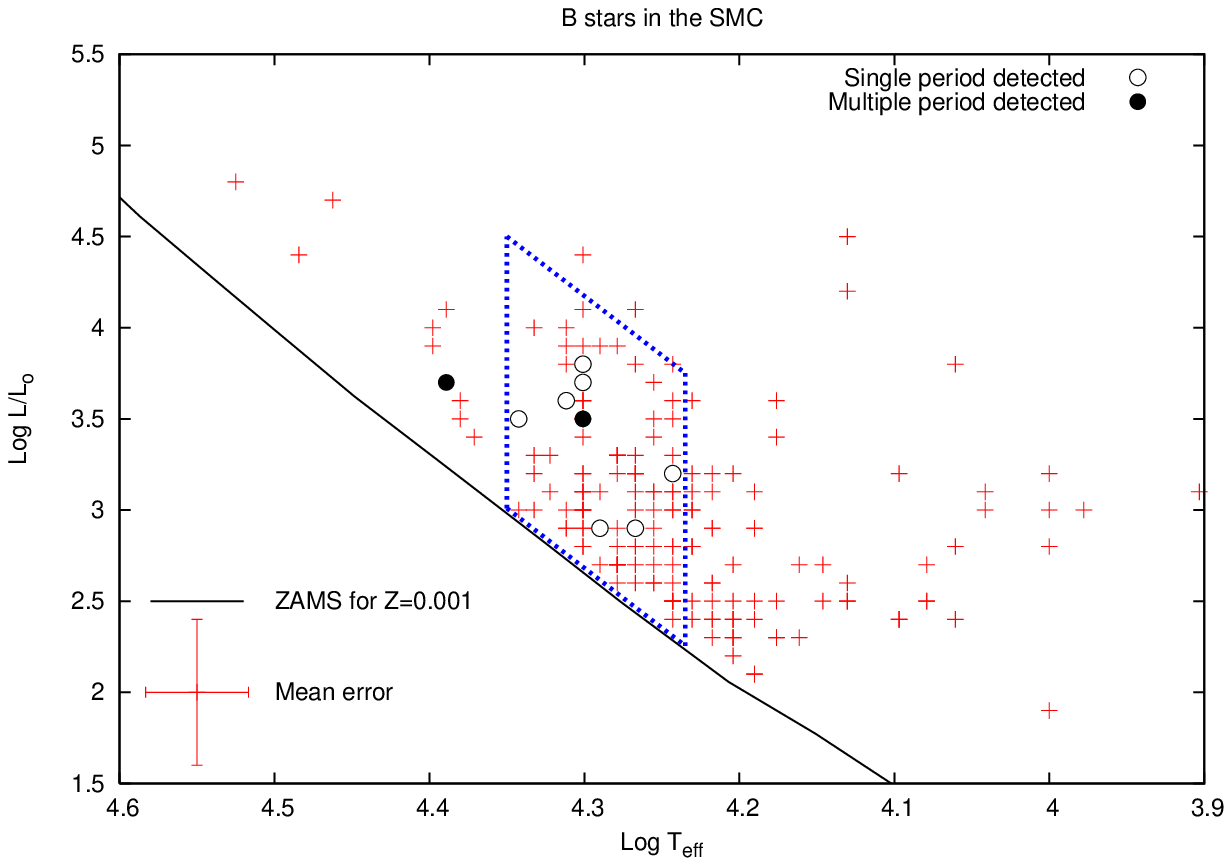}}
\caption{Location of the B star sample in the theoretical HR diagram. The dashed line delimits the region that contains all SPB stars found in this work.}
\label{H-RB-INEST}
\end{figure}

\begin{figure}
\centering
\resizebox{0.9\hsize}{!}{\includegraphics{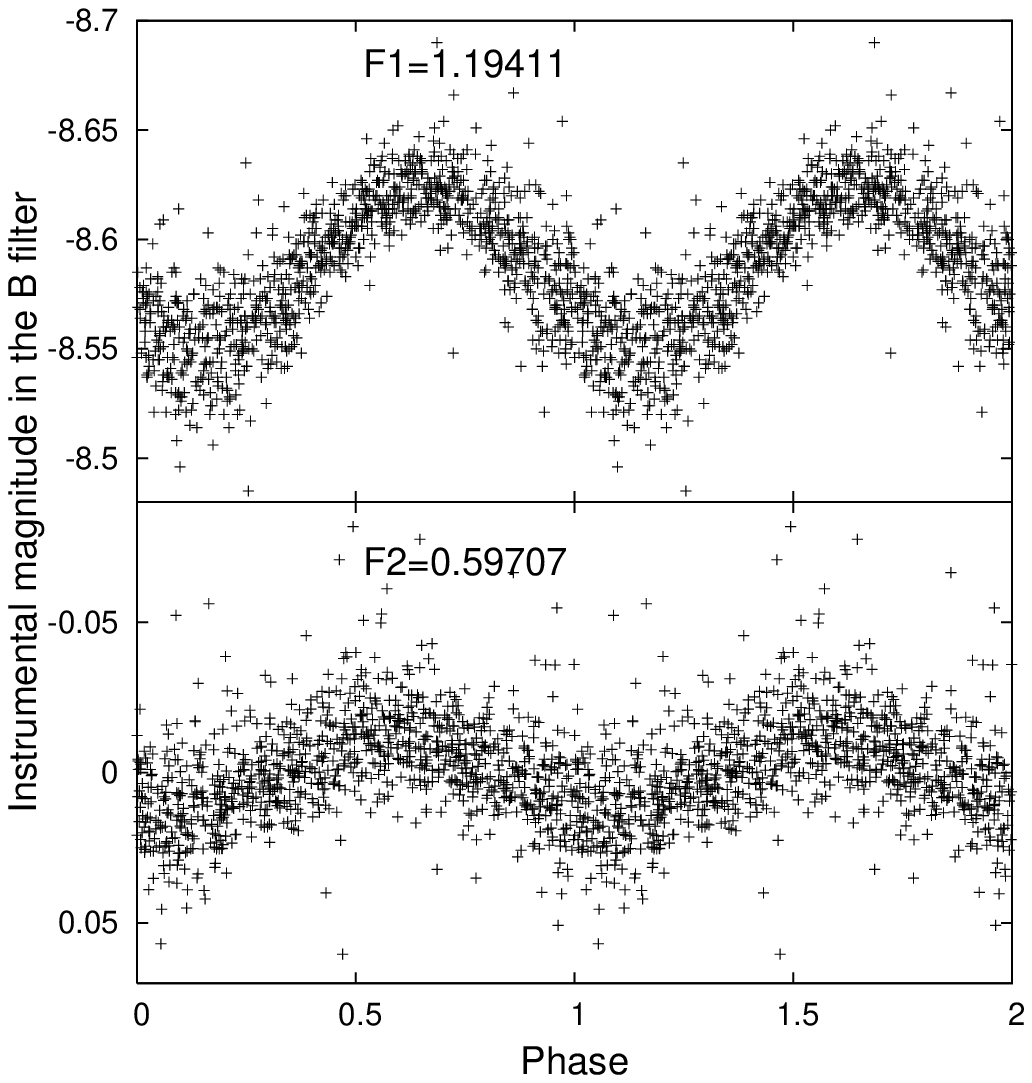}}
\caption{Phase diagram of the SPB star SMC5\_050662 folded with the two detected frequencies.}
\label{fig:phase:SPB}
\end{figure}

The eight remaining pulsating stars in our sample are SPB variables. They are located on the main sequence of the HR diagram, in the temperature range between $\log T_{\mathrm{eff}}=4.24$ and 4.35~K (see Fig.~\ref{H-RB-INEST}). The region containing all the SPB stars is depicted with a dashed line.
The high temperature limit is shifted towards hotter temperatures with respect to the instability strip at solar metallicity, represented in Fig.~\ref{H-RB}. It is also placed at temperatures higher than the ones predicted by the new computations by \cite{2007CoAst.151...48M} for $Z=0.005$. The low-temperature limit is most likely of observational origin, due 
to the fact that there are few stars with lower temperatures in our sample, and their photometry is less accurate because of their faintness. In Fig.~\ref{fig:phase:SPB} we display an example of the phase diagrams corresponding to 
a multiperiodic SPB star.
A puzzling circumstance regarding our sample of SPB stars is that only one has been detected as multiperiodic. The only multiperiodic object has a 
secondary period which is exactly twice the primary one, and hence even this 
object could be monoperiodic but with a non-sinusoidal light curve. This means 
that the variability in some of these stars might not be caused by pulsations, but
by other phenomena like eclipsing or ellipsoidal binarity. However, 
none of these stars have been detected as binary in the spectroscopic survey
of \citet{Martayan2007}. Thus, we consider our figure of eight stars as an upper limit 
to the number of bonna-fide SPB stars in our sample.

SPB stars in the Galaxy are also characterized as having low-rotational velocities, although there are several
cases of fast rotation \citep{2002ASPC..259..196D}. In our sample, SPB stars in the SMC have rotational velocities
evenly distributed around the mean rotational velocity of absorption-line B stars ($V\!\sin i=160$ km~s$^{-1}$,
M07).

\subsection{Pulsating Be stars}

\begin{figure}
\resizebox{\hsize}{!}{\includegraphics[bb=50 50 410 284,clip]{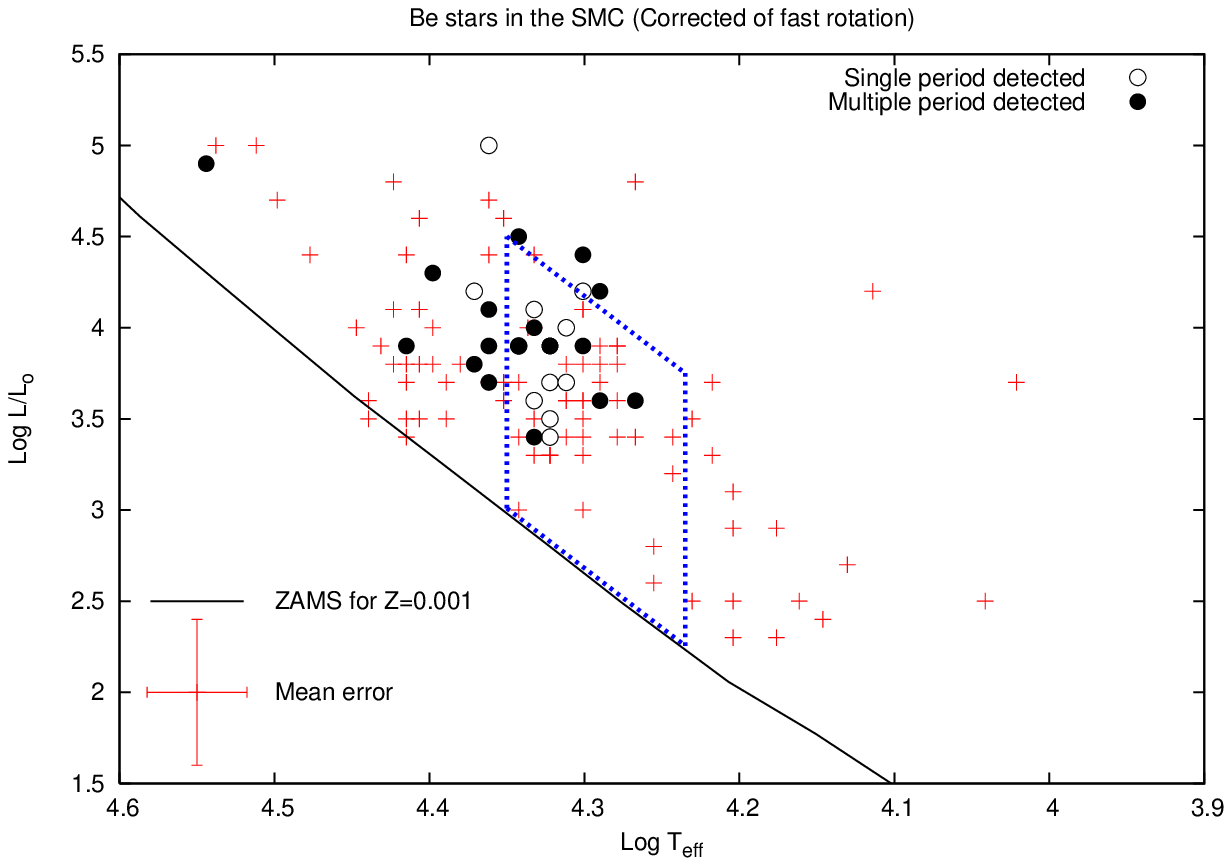}}
\caption{Location of the Be stars in the theoretical HR diagram. The dashed line represents the suggested 
SPB instability strip for the SMC.}
\label{H-RBe-INEST}
\end{figure}

In Fig.~\ref{H-RBe-INEST} we have represented the pulsating Be stars in the theoretical HR diagram. The values of the fundamental parameters of the Be stars are corrected for rapid rotation assuming $\Omega/\Omega_{c}=95\%$ (M07). 
We have included in the figure the region where the SPB stars are located as 
described in the previous subsection. Most Be stars are placed inside or very close to this region, suggesting that most pulsating Be stars in the SMC are g-mode SPB-like pulsators. Three stars are significantly outside the strip towards higher temperatures (SMC5\_014271, SMC5\_014878 and SMC5\_037158). All of them are multiperiodic, with the detected periods lower than 0.3 days. Therefore, we propose that these stars may be $\beta$ Cephei-like pulsators. 

\begin{figure}
\centering
\resizebox{0.9\hsize}{!}{\includegraphics{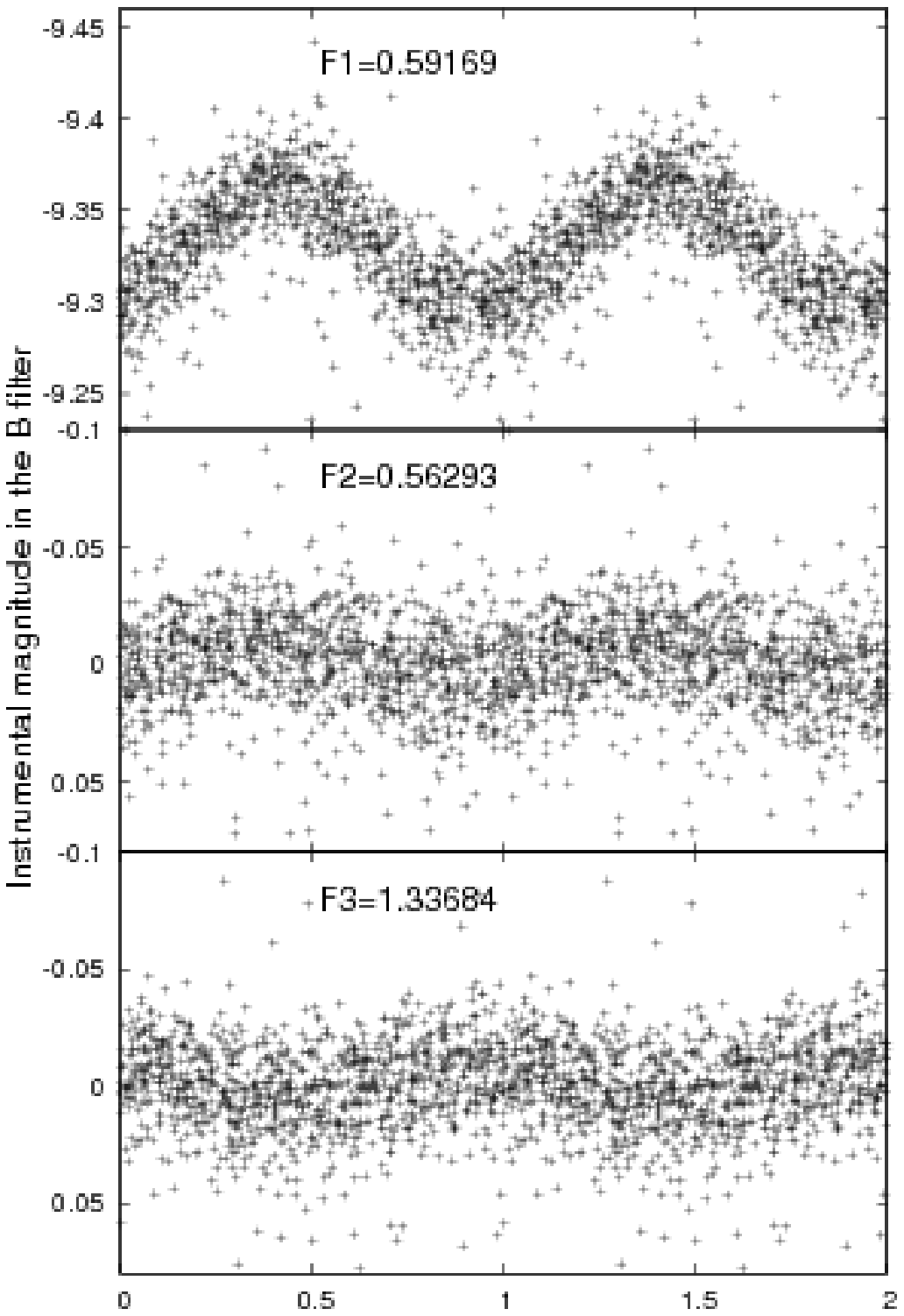}}
\caption{Phase diagram of the multiperiodic Be star SMC5\_045353 folded with the three detected frequencies.}
\label{fig:phase:Be}
\end{figure}

As an example, we display in Fig.~\ref{fig:phase:Be} the phase diagrams of the Be star SMC5\_045353, folded 
with the detected frequencies.

\subsection{Degree of variability}

The stellar samples searched for short-term variability are composed of 183 absorption-line B stars and 126 Be stars. We have found 9 pulsating stars among the former and 32 among the latter. This means that 4.9\% of B stars and 25.3\% of Be stars are short-term variables. If we restrict our comparison to the interval B2-B3, where most of the pulsating 
stars are found (see discussion below), the fraction of pulsating B and Be stars is 5.4\% and 31.5\% respectively.

The fraction of non-radial pulsators among the fast-rotating Be stars is much higher than the fraction among slow-rotating B stars. The same result was obtained for stars in the Galaxy by GS07. These results suggest that 
high rotational velocity has the effect of triggering the development of non-radial pulsations in B stars, or to enhance the amplitude of existing modes to make them more easily detectable. As an alternative explanation, 
the prevalence of non-radial pulsations could be related to the yet unknown nature of the Be phenomenon.


To compare the degree of variability between B stars in the SMC and in the Galaxy we make use of the 
statistics of pulsating B stars presented by GS07. We have only found one $\beta$ Cephei 
candidate in the SMC, and hence no statistical comparison is possible. 
In the case of SPB stars, in an unbiased sample of 795 galactic bright B stars,
GS07 found 35 SPB stars, ie. 4.4\%. In the SMC we found
8 SPB stars among 183 stars, representing a fraction of 4.4\%. 
Both percentages
are identical, and both samples are statistically significant. 

However, the above statistics are likely to be biased because in the galactic sample used by GS07, 
most of the stars are of late B-type, while in our SMC sample most of the stars have spectral types 
B3 or earlier. This is due to the fact that the late B stars are much fainter and are difficult to observe at the SMC distance. On the other hand, SPB stars are of spectral types B2 or later, both in the Galaxy and in the SMC.
If we restrict our comparison to the B2-B3 spectral range, in the Galaxy we have 14 SPB stars
among 160 B stars, ie. 8.8\%. In the SMC we have 8 SPB among 147 stars, representing 5.4\%.
Moreover, the much larger dataset of the MACHO survey allows for a 
much more complete detection than the Hipparcos data, from which most of
the known galactic SPB stars have been detected. Finally, the figure of eight 
SPB stars we have found in the SMC is to be considered as an upper limit 
as discussed in the previous section.
This represents a lower fraction in the SMC than in the Galaxy as predicted by
the current models.

All pulsating Be stars found in this work but one are of spectral types B3 or earlier. In fact, 
only 11 Be stars of the whole sample have later spectral types. The fraction of pulsating
Be stars in the SMC is 27\% in the B0-B3 interval. This percentage is to be compared with 
74\% found by GS07 or 86\% given by \citet{1998A&A...335..565H} for pulsating Be stars in the Galaxy.
Therefore, in the case of Be stars, the prevalence of pulsations in the SMC is significantly lower than in the 
Galaxy, as expected from the lower metallicity environment.

\section{Conclusions}

We have searched for short-term variability in a sample of 183 absorption-line B stars and 126 Be stars in the SMC with accurately determined physical parameters. We have studied their position in the theoretical HR diagram and mapped 
the regions of pulsational instability in the SMC.

We have found 9 pulsating absorption-line B stars. Among them, 8 have periods longer than 0.5 days, characteristic 
of Slow Pulsating B-type stars. The region occupied by the SPB stars in the SMC is shifted towards higher temperatures
with respect to the galactic SPB instability strip and to the predictions of recent models for $Z=0.005$. 
The remaining star presents two short periods within the range of the $\beta$ Cephei variables. This fact and 
its high effective temperature lead us to propose that it is indeed a $\beta$ Cephei star. This is 
an interesting result as current stellar models do not predict $\beta$ Cephei pulsations at the SMC metallicity.

There are 32 pulsating stars among the Be star sample. Most of them are placed in the region occupied by 
the SPB stars in the HR diagram of the SMC and present periods longer than 0.5 days. They are most probably 
g-mode SPB-like pulsators. 3 pulsating Be stars are significantly hotter and present much shorter periods, 
within the range of the $\beta$ Cephei stars. We propose that they are $\beta$ Cephei-like pulsators. 

The prevalence of pulsations among Be stars is significantly higher than among absorption-line B stars, much as 
in the Galaxy. We have found that 25.3\% of Be stars present short-term variability, while only 4.9\% of 
B star pulsates. This result indicates that the high rotational velocity either contributes to trigger the development of non-radial pulsations or to enhance the amplitude of the existing modes. Alternatively, 
the prevalence of non-radial pulsations could be related to the still unknown nature of the Be phenomenon.

Both the fraction of SPB among B stars and pulsating Be stars among the whole sample of Be stars in the SMC are lower than in the Galaxy, indicating that the prevalence of pulsations 
is directly affected by the metallicity of the environment, as predicted by
the current stellar models.

\begin{acknowledgements}
We thank the referee, Dr. Christoffel Waelkens, for his very useful comments.
This research has been financed by the Spanish ``Plan Nacional de Investigaci\'on Cient\'{\i}fica, Desarrollo e
Innovaci\'on Tecnol\'ogica'', and FEDER, through contract ESP~2004-03855-C03.
The work of P.~D. Diago is supported by a FPU grant from the Spanish ``Ministerio de Educaci\'{o}n y Ciencia''.
This paper utilizes public domain data obtained by the MACHO Project, jointly funded by the US Department of Energy through the University of California, Lawrence Livermore National Laboratory under contract No. W-7405-Eng-48, by the National Science Foundation through the Center for Particle Astrophysics of the University of California under cooperative agreement AST-8809616, and by the Mount Stromlo and Siding Spring Observatory, part of the Australian National University. 
\end{acknowledgements}

\bibliographystyle{aa}
\bibliography{8754}

\end{document}